%% file: ALpaper_arXiv_v2.tex
\documentclass[prb,showpacs,showkeys,twocolumn,amsmath,amssymb,amsfonts]{revtex4}
\pdfoutput=1
\usepackage{relsize}
\usepackage{graphicx}
\usepackage[usenames,dvipsnames]{color}
\usepackage[pdftex,colorlinks=true,linkcolor=Brown,citecolor=Brown,hyperfootnotes,pdfpagelabels,urlcolor=Brown]{hyperref}
\newcommand{\bra}[1]{\left\langle #1\right|}              
\newcommand{\ket}[1]{\left| #1\right\rangle}              
\newcommand{\ev}[1]{\left\langle #1\right\rangle}        
\newcommand{\mv}[2]{\langle#1|#2\rangle}                  
\newcommand{\ipop}[3]{\left\langle#1\left|#2\right|#3\right\rangle} 
\newcommand{\abs}[1]{\left| #1\right|}                    

\newcommand{\brc}[1]{{\relsize{-1}(#1)}}

\begin{document}
\title{Induced Delocalization by Correlation and Interaction\\in the one-dimensional Anderson Model}
\pacs{61.43.-j, 03.75.Lm, 72.15.Rn}
\keywords{Anderson localization; disordered Bose--Hubbard model; ultra-cold atoms}
\date{\today}
\author{Conrad Albrecht}
\email{c.albrecht@thphys.uni-heidelberg.de}
\author{Sandro Wimberger}
\affiliation{Institut f\"ur Theoretische Physik\\Universit\"at Heidelberg\\Philosophenweg 19, D-69120 Heidelberg}
\begin{abstract}
\noindent We consider long-range correlated disorder and mutual interacting particles
according to a dipole--dipole coupling as modifications to the one-dimensional Anderson
model. Technically we rely on the \brc{numerical} exact diagonalization of the system's
Hamilitonian. From the perspective of different localization measures we confirm and extend
the picture of the emergence of delocalized states with increasing correlations. Besides
these studies a definition for multi-particle localization is proposed. In the
case of two interacting bosons we observe a sensitivity of localization with respect
to the \textit{range} of the particle--particle interaction and insensitivity to
the coupling's sign, which should stimulate new theoretical approaches and experimental
investigations with e.g.\ \textit{dipolar} cold quantum gases.
\end{abstract}
\maketitle

\section{Introduction \& Motivation}
\label{sec:IntroMot}
\noindent When P.\ W.\ Anderson introduced a simple quantum model to represent a disordered lattice
\cite{Ande58PhysRev} it turned out that the contained physics is surprisingly complex, i.e.\
there exists the phenomenon of \textsl{Anderson localization} which is related
to an exponential decay of the quantum mechanical probability distribution. A specific
subclass of Hamiltonians Anderson studied reads
\begin{equation}
H_s=\sum^L_{i=1}{\epsilon_i c^+_i c_i}+J\sum^{L-1}_{i=1}{c^+_i c_{i+1}} + h.c.\quad,
\label{eq:AndHamSP}
\end{equation}
which represents a single particle in a one-dimensional chain of
$L$ sites with random \brc{\textit{onsite}} potential $\epsilon_i$ and kinetic energy $J$
\brc{\textit{hopping energy}}. We assume the formalism of second quantization
where the operator $c_i^{(+)}$ annihilates \brc{creates} a boson at site $i$ and
hence we refer to \hyperref[eq:AndHamSP]{Eq.\ (\ref{eq:AndHamSP})} as the \textit{disordered
Bose--Hubbard model} \brc{without mutual interaction}.
One can either derive $H_s$ from the \textit{tight binding approximation} of a continuous
model \cite{Altl10CMFT7.2} of non-interacting particles in an external
potential or one ab initio takes it as a discrete model.

Based on the renormalization group flow\footnote{As usual in condensed matter physics, the
\textit{flow} is parametrized by the system size.} idea
one can argue that the conductance of a disordered solid may vanish for sufficient large
systems \cite{Abra79PhysRevLett}. In particular the one-dimensional disordered system,
\hyperref[eq:AndHamSP]{Eq.\ (\ref{eq:AndHamSP})}, becomes an insulator in the limit $L\to\infty$.

An explicit argument for localization of all states $\ket{E}$ satisfying $H_s\ket{E}=E\ket{E}$
can be established by exploiting the \textit{transfer matrix method} \cite{Haak01qsc}
and a theorem due to F\"urstenberg \cite{Furs63NonCommRM}.
On the other hand there is Bloch's theorem \cite{Bloc28PhysRev} which
induces periodic \brc{delocalized} states for a periodic lattice potential $\epsilon_i$.
Hence the random nature of the potential must be obviously the key feature that leads
to localization: A criterion based on the \textit{differentiability} of the disorder
potential $\epsilon_i$ was recently studied to understand the \textit{degree of randomness}
necessary for delocalization \cite{Garc09PhysRevB}.

In order to investigate the impact of correlation on localization we introduce a specific disorder model
that extrapolates from a \textit{pure} random sequence $\epsilon_1,\epsilon_2,\dots, \epsilon_L$ to a periodic, and thus
correlated, structure in \hyperref[sec:CorrDis]{Sect.\ \ref{sec:CorrDis}}. It was first used by Moura \& Lyra \cite{Mour98Deloc1DAMCorrDis} and
similar investigations followed \cite{Schu05PhysicaB,Kaya07EurPhysB}. In a first step we
will review the model on the basis of three different localization measures. Furthermore,
we utilize one of those quantities to establish the \textit{phase diagram} --- namely, the dependence
of localization with respect to disorder strength and the amount of correlation among the $\epsilon_i$. While
a previous study by Shima et al.\cite{Shim04PhysRevB}\ focused on the properties of states in the band center,
our measure in use accounts for the \textit{global} aspect, i.e.\ it incorporates properties of
the entire spectrum.

The potential interest in the sensitivity of Anderson localization on correlated disorder arose due to recent
experiments with Bose--Einstein condensates \brc{BECs} where the direct observation
of the atomic density distribution provides access to the quantum mechanical probability
distribution \cite{Bill08nat,Pale10natphys}. Besides correlation the impact of interaction
is an issue one is naturally faced with when studying localization in BECs.
Anderson already mentioned the importance of particle interaction \cite{Ande78NoblLect} on
localization and worked out theoretical investigations in collaboration with L.\ Fleishman \cite{Flei80PhysRevB}.
Over the years various aspects and features of interacting particles in a random potential
have been figured out, but the problem remains a challenging topic for present research
since results from different approaches do not always coincide. Just recently,
the phase diagram of the 3-dimensional disordered Bose--Hubbard model was established \cite{Gura09PhysRevB}.

In our discussion on interacting particles in the presence of a disordered onsite potential
\brc{\hyperref[sec:LocInt]{Sect.\ \ref{sec:LocInt}}} we first want to focus on a suitable
definition of multi-particle localization and then turn to the extension of
\hyperref[eq:AndHamSP]{Eq.\ (\ref{eq:AndHamSP})},
namely
\begin{equation}
H_{mp}=H_s+\sum_{i,j=1}^{L}U_{ij}c^+_ic^+_jc_ic_j\quad,
\label{eq:AndHamMP}
\end{equation}
with a \brc{two-body} interaction potential $U_{ij}$. We model interaction according to a magnetic dipole--dipole
coupling obtained in BEC experiments with \textit{dipolar gases} \cite{Grie05PhysRevLett,Stuh05PhysRevLett}
which is beyond the standard treatment of the \textit{onsite interaction term} $U_0\sum_i \hat{n}_i(\hat{n}_i-1)$
\brc{with $\hat{n}_i\equiv c^+_ic_i$} present in the disordered Bose--Hubbard model. By explicitely
diagonalizing $H_{mp}$ for two interacting bosons we explore the relevance of interaction \brc{especially
its \textit{range}} for localization and discuss an interesting symmetry involving the \textit{sign} of the interaction
potential $U_{ij}$.

\section{Correlated Disorder}
\label{sec:CorrDis}
\noindent To model correlated disorder we use the following prescription for the onsite potential
values $\epsilon_i$ in \hyperref[eq:AndHamSP]{Eq.\ (\ref{eq:AndHamSP})}:
\begin{equation}
\epsilon_i=\sum_{k=1}^{N/2}{\left[\frac{2\pi k}{N}\right]^{-\alpha/2}
\cos{\left(\frac{2\pi}{N}ki+\phi_k\right)}}\quad{\small\begin{cases}&\alpha\geq0\\&N\gg L\end{cases}}\quad.
\label{eq:ProdCorrDis}
\end{equation}
Here, $\alpha$ denotes the \textit{correlation parameter}, $\phi_k\in[0,2\pi)$ are
$\frac{N}{2}$ uniformly distributed random phases and $N$ is a natural number that
should be much larger than the number of lattice sites\footnote{In
our numerics with $L\sim10^3-10^4$ we choose $N_0=10^5$ and ensure that our results are
not altered for $N>N_0$ by means of finite size analysis, i.e.\ we do not monitor
any qualitative deviation of the data for $N>N_0$.} $L$. Roughly speaking, the $\epsilon_i$
\brc{$i=1\dots L$} represent the discrete Fourier transform of $k^{-\alpha}$, i.e.\ an algebraic decaying
\textit{power spectrum} \cite{Osbo88PhysicaD}. One is therefore used to refer to the
$\epsilon_i$ as \textit{long-range} correlated and indeed if we turn back to the
continuum limit, we can argue that the $\epsilon_j$ are correlated according to an
algebraic decay for $\alpha\in(0,1)$. The case $\alpha=0$ corresponds to \textit{almost}
uncorrelated disorder which is close to the \textit{perfect} disorder Anderson assumed in his model.
More details on properties of the $\epsilon_i$ from \hyperref[eq:ProdCorrDis]{Eq.\ (\ref{eq:ProdCorrDis})}
are provided in \hyperref[app:CorrDis]{App.\ \ref{app:CorrDis}}.

\subsection{Localization Measures}
\label{sec:LocMeas}
\noindent In order to detect the localization--delocalization transition we introduce
three different measures:\\
a) the \textit{Normalized Standard Deviation} \cite{Moor73JPhysC} \brc{NSD}\\
b) the \textit{Inverse Participation Ratio} \cite{Kram93RepProgPhys} \brc{IPR} and\\
c) the \textit{Nearest Neighbor Distribution} \cite{Brod81RevModPhys} \brc{NND}.\\
While the NSD and the IPR are derived from the \textit{spatial/site} probability distribution
\begin{equation}
\psi^2_E(i)\equiv\abs{\mv{i}{E}}^2\text{with}~H_s\ket{E}=E\ket{E}~\text{and}~\ket{i}\equiv c^+_i\ket{0},
\label{eq:DefNotProbDist}
\end{equation}
where $c_i\ket{0}=0$, the NND depends on the spectral values $E$ only. Given a chain of $L$ sites we calculate
\begin{equation}
\text{NSD}(E)\equiv\frac{\ev{i^2}-\ev{i}^2}{(L^2-1)/12}
\label{eq:NSD}
\end{equation}
where $\ev{.}\equiv\ipop{E}{.}{E}$ denotes the expectation value and hence the NSD
derives the \brc{spatial} variance of the lattice site index $i$ for some $\ket{E}$
with respect to the state $\ket{\psi}$ whose probability distribution is uniform
on the lattice.

In the case of the IPR one quantifies the \brc{inverse} number of
sites where $\psi^2_E(i)$ \textit{significantly} differs from zero. Since $\sum_{i=1}^L\psi^2_E(i)=1$
we state that
\begin{equation}
\text{IPR}(E)\equiv\sum_{i=1}^L\left[\psi^2_E(i)\right]^2\sim\frac{1}{\tilde{L}}
\label{eq:IPR}
\end{equation}
with $\tilde{L}$ defined as the number of sites that are \textit{occupied} by $\ket{E}$. Indeed, we
can convince ourselves that $\text{IPR}\sim1$ and $\text{IPR}\sim L^{-1}\xrightarrow{L\to\infty}0$
for localized and delocalized states, respectively.

In contrast, the NND considers the spectral properties of $H_s$. More precisely, we
evaluate fluctuations of level spacings around a \textit{local} mean $\bar{s}_n$
by computing
\begin{equation}
s_n\equiv(E_{n+1}-E_n)/\bar{s}_n~\text{with}~\bar{s}_n\equiv\frac{E_{n+1+\frac{m}{2}}-E_{n-\frac{m}{2}}}{m+1},
\label{eq:NNDDef}
\end{equation}
and deriving the distribution $P(s)$ of the \textit{nearest neighbor spacings} $s_n$.
Here, we labeled the spectral values according to $E_1\leq E_2\leq\dots\leq E_L$
and the division by $\bar{s}_n$ \textit{unfolds} \cite{Haak01qsc}
the level spacings to relate different parts of the whole spectrum to each other. The
procedure of unfolding is not unique \cite{Gome02PhysRev} and $m$ is left as a free
parameter that defines the notion of \textit{local}. We choose it such that $m\ll L$
on the one hand and to include enough energy values for reasonable statistics on the
other hand; in fact, we used $m\sim 20$ for $L\sim10^3$.

The NND is not as obvious as the former measures \cite{Efet04RMTAndSUSY,Simo96Univers}. Intuitively,
the argument works as follows: Taking two different energy eigenstates $\ket{E},\ket{E'}$
that do not significantly \textit{overlap} \brc{in site space} we assume them to be \textit{almost orthogonal},
i.e.\ they are in some sense \textit{independent} from each other \brc{localized}
and nothing prevents the corresponding energy values to be arbitrary close \brc{$s\to0$}. But if the overlap
increases \brc{extended states} the levels start to \textit{repell}\footnote{On the basis of coupling
two harmonic oscillators the notion of \textit{level repulsion} becomes a bit more explicit. In fact $H_s$
from \hyperref[eq:AndHamSP]{Eq.\ (\ref{eq:AndHamSP})} just mimics such a system with ground
state energy difference $\Delta\epsilon\equiv\abs{\epsilon_1-\epsilon_2}$ and \textit{\brc{ground state}
coupling energy} $J$ between the two oscillators labeled by $i=1,2$ in the most
simplest case of $L=2$. By investigating the quantum physics of a single particle in such a setup
with $\epsilon_1=0$ one observes that the energy difference of the system's eigenstates reads
$\Delta E\equiv\abs{E_1-E_2}=2\sqrt{\Delta\epsilon^2/4+J^2}$, i.e.\ in the uncoupled case
$J=0$ the eigenstates $\ket{E_1=0}$ and $\ket{E_2=\Delta\epsilon}$ are localized in one of the harmonic
oscillators and by tuning the external parameter $\Delta\epsilon\to0$ their corresponding eigenenergies
$E_i$ become degenerate. By means of $\Delta E$ this is impossible for $J\neq0$ where both
eigenstates $\ket{E_1},\ket{E_2}$ are \textit{distributed} among the two coupled oscillators.}: $E\neq E'$.
Quantities that are statistically independent exhibit a Poissonian distribution and
thus the corresponding NND should be \cite{Fish85PhysRevB} $P(s)\sim e^{-s}$. In contrast,
$P(0)=0$ is reasonable to expect for delocalized states.

\subsection{Numerical Results}
\label{sec:CorrDis_NumRes}
\begin{figure}[t!]
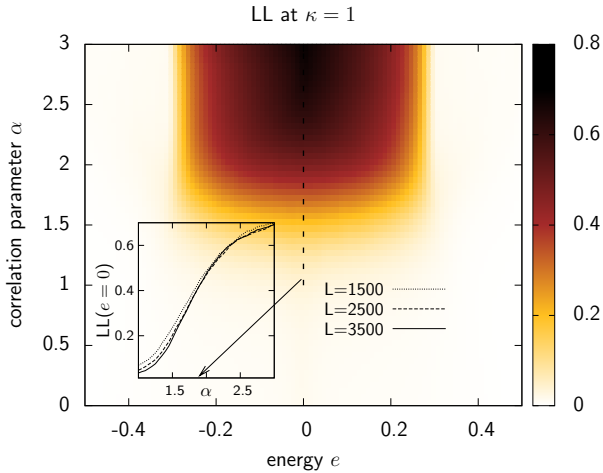

\centering
\resizebox{0.48\textwidth}{!}{\relsize{+2}\input FIG1}
\caption{\label{fig:LLComp}\brc{color online} Energy--resolved plot of the \textit{localization length}
\brc{LL} for different correlation parameter values $\alpha$ to compare our results
from exact \brc{numerical} diagonalization to investigations that use the transfer matrix method
to obtain the Lyapunov exponent $\gamma$ \brc{cf.\ Ref.\ [\onlinecite{Mour98Deloc1DAMCorrDis}],
fig.\ 5}. We used a system of $L=3500$ sites and averaged over $50$ disorder realizations.
The inset \brc{transparent to the main panel in the background} presents numerical data of $\text{LL(0)}$
for different system sizes which underpins the reported delocalization transition around $\alpha=2$.}
\end{figure}
\noindent Turning back to \hyperref[eq:AndHamSP]{Eq.\ (\ref{eq:AndHamSP})} we rescale $H_s$ by $J$,
i.e.\ $H_s\to J^{-1}H_s$, which does not alter the Hamiltonian's eigenstates but 
multiplies the spectral values $E$ by a factor of $J^{-1}$. Since we restrict the $\epsilon_i$ to
the \brc{finite} interval $[-\frac{\Delta}{2},\frac{\Delta}{2}]$
we introduce the parameter
\begin{equation}
\kappa\equiv\Delta/J
\end{equation}
that indicates the \textit{strength}
of disorder. Thus we are faced with a two--dimensional set of system parameters
$(\alpha,\kappa)$ and, from the following as well as \hyperref[app:CorrDis]{App.\ \ref{app:CorrDis}},
it becomes clear that increasing $\alpha$ corresponds to increasing correlations
up to \textit{long-range} correlated disorder. Moreover, since the NSD and the IPR
only depend on $\ket{E}$, we rescale and shift the spectrum such that the
rescaled values $E\to e$ obey $-0.5\leq e\leq0.5$ when plotting these quantities
resolved in energy.

To set the stage, we want to relate our numerical results to previously published ones.
Since it is common practise to attempt localization by computing the \textit{Lyapunov exponent}\footnote{Despite
similar in notion the Lyapunov exponent $\gamma$ in localization theory has not to
be confused with its counterpart in nonlinear dynamics \cite{Stro94NonDynChaosCh10.5}
$\gamma$ \brc{where it is often denoted as $\lambda$}. Here, $\gamma$ is equivalent to the
exponential decay rate of $\psi^2_E$ \brc{cf.\ \hyperref[eq:DefNotProbDist]{Eq.\ (\ref{eq:DefNotProbDist})}}
for sufficient large systems.}~~\cite{Haak01qsc},
we establish an exponential fit to $\ket{E}$. More precisely, our numerics picks out the
maximum $\psi^2_E(i_0)$ and performs an exponential fit into the \textit{direction} singled
out by $\max(i_0,L-i_0)$. From $\psi^2_{E,fit}(i)\sim\exp{(-\gamma\abs{i-i_0}})$ we
extract $\gamma(E)$ and define the \textit{localization length}
\begin{equation}
\text{LL}(E)\equiv\gamma^{-1}(E)/L\quad.
\end{equation}
The corresponding result for \textit{intermediate} disorder \brc{$\kappa=1$} is shown in
\hyperref[fig:LLComp]{Fig.\ \ref{fig:LLComp}} and it exhibits reasonable qualitative
agreement with Ref.\ [\onlinecite{Mour98Deloc1DAMCorrDis}],
fig.\ 5; namely, delocalized states \brc{$\gamma^{-1}\sim L\rightarrow\text{LL}\sim1$} arise
in a finite range around the \textit{band center} $e=0$ when the correlation within the
disorder potential $\epsilon_i$ is increased. This interpretation also coincides
with the plot presented in Ref.\ [\onlinecite{Schu05PhysicaB}], fig.\ 3, where localization
was quantified by means of a measure based on the \textit{density of states}
\begin{equation}
\text{DOS}(E)\equiv \frac{dn}{dE}
\label{eq:DefDOS}
\end{equation}
which defines the number $dn$ of energy eigenstates $\ket{E}$ in a given interval $[E,E+dE]$\footnote{In
particular $\int_{E_1}^{E_2}dE~\text{DOS}(E)$ is the number of states $\ket{E}$ with $E_1\leq E\leq E_2$
and rescaling the DOS simultaneously with the spectrum $\{E\}$ is understood to fulfill
the normalization condition $1=\int dE \text{DOS}(E)$.}.

Moreover, the work of Moura \& Lyra mentioned above supports delocalization at $\alpha=2$ for states at the
\begin{figure}[t!]
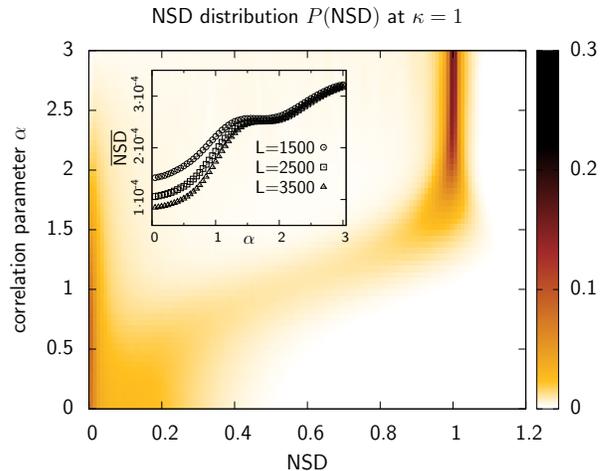

\centering
\resizebox{0.48\textwidth}{!}{\relsize{+2}\input FIG2}
\caption{\label{fig:NSDCorrDis}\brc{color online} Plot of the distribution $P(\text{NSD})$
of the \brc{spatially} normalized standard deviation \hyperref[eq:NSD]{Eq.\ (\ref{eq:NSD})}
at \textit{intermediate  disorder} $\kappa=1$ \brc{$J=\Delta$} to support that localization$\to$delocalization 
takes place rather smoothly than according to a sharp transition. As for \hyperref[fig:LLComp]{Fig.\ \ref{fig:LLComp}}
we used $L=3500$ and the averaging was taken over $50$ disorder realizations. The inset \brc{transparent to its
background} shows results for different system sizes and supports our crossover picture as well \brc{details in the main text}.}
\end{figure}
band center. Therefore we depicted a cut of \hyperref[fig:LLComp]{Fig.\ \ref{fig:LLComp}} at $e=0$ in the
relevant correlation parameter range $\alpha\in[1,3]$ and checked $\text{LL}(0)$ according to a finite size analysis,
inset of \hyperref[fig:LLComp]{Fig.\ \ref{fig:LLComp}}. Indeed, up to $\alpha\approx2$ the $\text{LL}(0)$ decreases
with increasing system size $L$ while above this correlation parameter value it remains relatively constant.
Assuming that this trend persits for even larger system sizes, the data support that $\alpha=2$ marks a qualitative difference
between systems with smaller and larger correlation, respectively: While the relative extend of the states
with respect to the system size falls off \brc{localization} for $0\leq\alpha\lesssim2$, it remains constant
\brc{extended states} for $\alpha\gtrsim2$ in the \textit{thermodynamic limit} $L\to\infty$.

However, we would like to address the question \textit{where} \brc{in terms of $\alpha$} does the
localization--delocalization transition takes place from the perspective of the measures introduced in
\hyperref[sec:LocMeas]{Sect.\ \ref{sec:LocMeas}}: Instead of a sharp transition at $\alpha=2$, which is
supported by \hyperref[fig:LLComp]{Fig.\ \ref{fig:LLComp}} and publications mentioned above, we suggest
a smooth crossover in $1\lesssim\alpha\lesssim2$. Investigating the distribution $P(\text{NSD})$
most obviously illustrates this statement and we present it in \hyperref[fig:NSDCorrDis]{Fig.\ \ref{fig:NSDCorrDis}}.
Although the major fraction of localized states \brc{$\text{NSD}\ll1$} becomes delocalized
at $\alpha\approx2$ there is a finite fraction that \textit{splits} apart the localized
region \brc{$0\leq\text{NSD}\lesssim0.2$} around $\alpha=0.5$ and \textit{drifts}
\begin{figure}[t!]
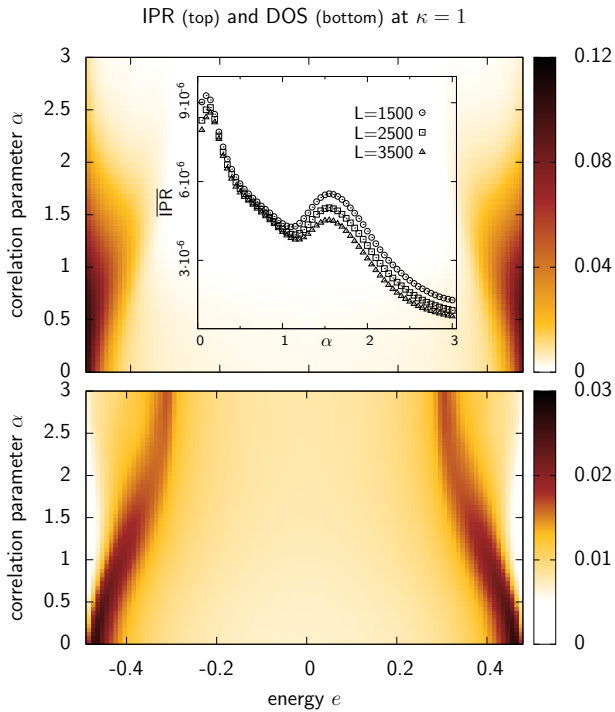

\centering
\resizebox{0.48\textwidth}{!}{\relsize{+2}\input FIG3a}
\resizebox{0.48\textwidth}{!}{\relsize{+2}\input FIG3b}
\caption{\label{fig:IPRRes}\brc{color online} Smooth localization--delocalization
crossover from the perspective of the simultaneous analysis of the IPR and the DOS.
Increasing correlation \textit{attracts} the system's states from the \textit{band
edges} \brc{$\abs{e}=0.5$} to the \textit{band center} \brc{$e=0$} where delocalized
states are present and thus the system becomes more and more delocalized. Finally, localized
states almost completely vanish above $\alpha=2$.
The inset of the upper panel \brc{transparent to the background again} establishes the finite
size analysis similar to the inset of \hyperref[fig:NSDCorrDis]{Fig.\ \ref{fig:NSDCorrDis}}.}
\end{figure}
towards the delocalized regime \brc{$\text{NSD}\approx1$} up to $\alpha\approx2$. This observation remains
\textit{stable} in the numerically studied range of system sizes \brc{$L=1000\dots3500$}.

An explicit plot to the finite size analysis performed with respect to our crossover picture is given
by the inset of \hyperref[fig:NSDCorrDis]{Fig.\ \ref{fig:NSDCorrDis}}. Convoluting the NSD with the DOS, i.e.\
\begin{equation}
\overline{\text{NSD}}\equiv\int dE~\text{NSD}(E)\text{DOS}(E)
\label{eq:NSDAvDef}
\end{equation}
yields a single quantity to characterize the \textit{global} localization property of the disorderd system at fixed
correlation parameter $\alpha$. In the spirit of the previously applied
LL--dependence on $L$, the inset of \hyperref[fig:NSDCorrDis]{Fig.\ \ref{fig:NSDCorrDis}} provides the
system size dependence of the \textit{averaged normalized
standard deviation} $\overline{\text{NSD}}$: Up to $\alpha\approx1$, $\overline{\text{NSD}}$ decreases
with increasing system size $L$ indicating that the Hamiltonian's eigenstates stay localized in the
thermodynamic limit. Above $\alpha\approx1$ the averaged standard deviation that is normalized to
$L$ remains unaltered for the accessed number of sites --- a signature of delocalization. Remarkably, the
$\overline{\text{NSD}}$ forms a plateau in $1\lesssim\alpha\lesssim2$, before growing up in absolute
value with increasing correlation. One more evidence to the crossover picture we are in favor of.

Similarly, we investigate the IPR taking into account the DOS from \hyperref[eq:DefDOS]{Eq.\ (\ref{eq:DefDOS})}.
Since we are interested in the localization property of the whole system it is necessary
to account for the number of states with a certain localization measure value. Hence
the IPR and the DOS are plotted simultaneously in \hyperref[fig:IPRRes]{Fig.\ \ref{fig:IPRRes}}
and the inset of the upper panel presents an averaged version of the inverse participation ratio in total analogy to
\hyperref[eq:NSDAvDef]{Eq.\ (\ref{eq:NSDAvDef})} with $\text{NSD}(E)$ replaced by $\text{IPR}(E)$.

From a naive point of view we could conclude that even for weak correlation \brc{$\alpha\lesssim0.5$}
the system is delocalized since there are only a few localized states near the \textit{band edges} \brc{$\abs{e}=0.5$} while the
rest of the spectrum exhibits an IPR value corresponding to delocalization.
Moreover, increasing correlation enhances localization up to $\alpha\approx1$ which
seems unexpected. But if we include the DOS \brc{lower panel of \hyperref[fig:IPRRes]{Fig.\ \ref{fig:IPRRes}}} the
picture drawn before reveals: We realize that almost all states have energies near to the band edges for $\alpha\lesssim0.5$
and thus the system is localized. The crucial fact leading to the localization--delocalization
crossover is that for increasing correlation the band center \textit{attracts} the
states from the band edges to the regime where delocalized states are permanently
present \brc{$\abs{e}\lesssim0.3$}. Localized states at the band edges follow that
trend only up to $\alpha\approx1$ and for stronger correlation they start to disappear
until they are almost vanished at $\alpha\approx2$. According to an argument presented in
\hyperref[app:CorrDis]{App.\ \ref{app:CorrDis}} it is reasonable to refer to disorder with
$\alpha\geq1$ as \textit{highly correlated} and therefore it provides a hint to the
suggestive importance of $\alpha=1$ in our data.

Again, this observation is confirmed by a finite size analysis as shown in the inset of the upper
panel of \hyperref[fig:IPRRes]{Fig.\ \ref{fig:IPRRes}}. We note that the $\overline{\text{IPR}}$--values
stay quite unaltered for $\alpha\lesssim1$ when $L$ is increased: Since $\text{IPR}\sim\tilde{L}^{-1}\sim L^{-1}$ \brc{cf.\ \hyperref[eq:IPR]{Eq.\ \ref{eq:IPR}}}
corresponds to extended states which tends to zero in the limit $L\to\infty$, the decreasing behavior of $\overline{\text{IPR}}(L)$
for $\alpha\gtrsim1$ suggests the existence of delocalized states. The \textit{bump} in the
\textit{crossover interval} $\alpha\in[1,2]$ may represent a finite size effect that eventually vanishes
in the thermodynamic limit.

\begin{figure}[t!]
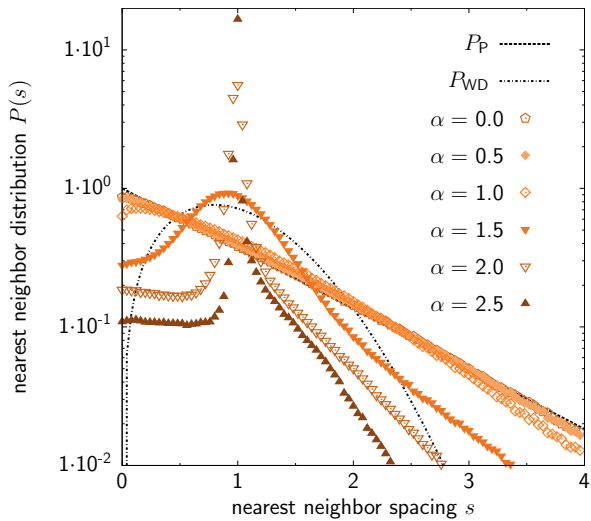

\centering
\resizebox{0.42\textwidth}{!}{\relsize{+3}\input FIG4}
\caption{\label{fig:NNDRes}\brc{color online} Semi-logarithmic plot of the nearest
neighbor spacing distributions for several correlation parameter values $\alpha$ \brc{$L=3\cdot10^4$, $\kappa=1$}.
The broken line is to compare them to the Poissonian distribution $P_\text{P}$
which we expect for localized states. As apparent, the distribution does not converge
to $P_\text{WD}$ of the \textit{Gaussian orthogonal ensemble}\cite{Haak01qsc}
since for $\alpha\gg1$ the onsite potential $\epsilon_i$ becomes periodic, local fluctuations among the $E_i$ vanish
and thus $P(s)$ must be peaked around $s=1$ \brc{$s_n=(E_{n+1}-E_n)/\bar{s}_n\approx1,~\alpha\gg1$}.
A more detailed discussion is given in the main text.}
\end{figure}

Finally, we would like to take a closer look at the smooth crossover from the perspective of the energy
spectrum. Therefore we investigate the NND from \hyperref[sec:LocMeas]{Sect.\ \ref{sec:LocMeas}}
which also prepares the relevant measure we will use to quantify the $(\kappa,\alpha)$--dependence
of localization. Plots of the NND for different correlation parameters $\alpha$ are
presented in \hyperref[fig:NNDRes]{Fig.\ \ref{fig:NNDRes}}. For small $\alpha$
the result is near to the Poissonian distribution $e^{-s}$. Deviations from the theoretical
prediction are due to the finite system size $L$ and the fact that $\alpha=0$ does
not exactly correspond to uncorrelated but weakly correlated disorder \brc{see
\hyperref[app:CorrDis]{App.\ \ref{app:CorrDis}}}.
We further note that the data points for $\alpha=0$ and $\alpha=0.5$ almost coincide,
but when increasing $\alpha$ above $0.5$ the distribution starts to deviate more
significantly from $e^{-s}$ to become peaked around $s=1$. Returning
to \hyperref[eq:ProdCorrDis]{Eq.\ (\ref{eq:ProdCorrDis})} we note that for $\alpha\gg1$ the
summation over the \brc{discrete} momenta $k$ \textit{picks up} less cosine modes
which leads to a periodic onsite potential in the limit $\alpha\to\infty$. Hence
the spectrum of $H_s$ becomes more regular for increasing $\alpha$ and \brc{local}
fluctutations eventually vanish for sufficiently strong correlation due to Bloch's theorem. From the definition
of the unfolded level spacings $s_n$, \hyperref[eq:NNDDef]{Eq.\ (\ref{eq:NNDDef})},
we conclude that all $s$ values should be centered around $1$ due to the normalization
by the \text{local mean} $\bar{s}_n$.

At this stage it is maybe advisable to discuss the \textit{Wigner surmise} $P_\text{WD}$ --- a
specific NND connected to \textit{random matrix theory}\cite{Haak01qsc} --- in the context of
localization, especially in our specific model system under consideration. In the literature\cite{Shkl93PhysRevB,Jacq98PhysRevLett,Cuev99PhysRevLett}
it seems to be common practise to associate 
\begin{equation}
P_\text{P}(s)\equiv\exp(-s)
\label{eq:LocDelCorrNND}
\end{equation}
with localized states and signatures of level repulsion \brc{$P(s\ll1)\ll1$} in the NND with extended ones after
having unfolded the spectrum of the corresponding disordered system. In the case of a
\textit{time reversal symmetric quantum chaotic system} --- represented by real, symmetric
random matrices --- the NND adopts the \textit{Wigner surmise}
\begin{equation}
P_\text{WD}(s)\equiv\frac{\pi}{2}s\exp\left(-\frac{\pi}{4}s^2\right)
\label{eq:DefWDDist}
\end{equation}
in the thermodynamic limit \brc{cf.\ Ref.\ \onlinecite{Haak01qsc} as well as references therein}.

As we scetched it in \hyperref[sec:LocMeas]{Sect.\ \ref{sec:LocMeas}}
it is quite plausible to adopt the first correspondence. Indeed, the matrix representation of $H_s$
in the site basis $\{\ket{i}\}$ becomes \textit{tridiagonal}, i.e.\
\begin{equation}
H_{s,ij}\equiv\ipop{i}{H_s}{j}
\end{equation}
vanishes except for its diagonal elements $H_{s,ii}=\epsilon_i$ and the first off-diagonal entries $H_{s,ii+1}=H_{s,i+1i}=J$
when \textit{open} boundary conditions \brc{cf.\ \hyperref[app:MatrRepMP]{App.\ \ref{app:MatrRepMP}}}
are imposed.  In the limit where the \brc{uncorrelated} disorder strongly dominates the kinetic energy
\brc{$\Delta\gg J$}, the single particle Hamiltonian becomes approximately diagonal to zeroth order in
$\tfrac{J}{\Delta}$ and the solution to the Schr\"odinger equation reads $\ket{E_i}\approx\ket{i}$
with $E_i\approx\epsilon_i$. Hence the almost purely random/uncorrelated sequence of eigenvalues $E_i$
should obey a Poissonian distribution in the NND.

However, there is a crucial fact we would like to keep in mind: Although the Hamilton matrix of
$H_s$ is real valued and symmetric \brc{in site space} --- since we deal with static disorder $\epsilon_i$ and
real valued constant hopping energy $J$ --- the elements $H_{s,ij}$ do not fulfill one of the conditions\footnote{As known
from standard text books, e.g.\ Ref.\ [\onlinecite{Haak01qsc}] or F.\ Dyson's work Ref.\ [\onlinecite{Dyso62JMathPhys}]
one assumes a) the joint probability $p(H_\text{rand})$ of the random matrix ensemble $H_\text{rand}$ being invariant
under orthogonal transformations and b) one requires the statistical independence of all Hamilton matrix elements
which does not hold in our case of $H_\text{rand}\leftrightarrow H_s$: Despite the random potential values $\epsilon_i$
the kinetic hopping energy $J$ is de\-ter\-mi\-nis\-tic/constant.} from which $P_\text{WD}$ is derived as the
corresponding NND of quantum chaotic systems. Therefore we will not naturally expect the Wigner surmise to arise in the presence of
delocalized states. Indeed, \hyperref[fig:NNDRes]{Fig.\ \ref{fig:NNDRes}} numerically verifies this conjecture.

An issue closely related to the analysis of the NND is the question on a \textit{critical
distribution} $P_c$ that marks the transition from localization to delocalization. One way to detect this transition
by means of the NND is to argue for a suitable intermediate distribution between $P_\text{P}$ and $P_\text{WD}$ as
done in e.g.\ Ref.\ [\onlinecite{Shkl93PhysRevB}]. Usually one investigates the quantity
\begin{equation}
\eta\equiv\frac{\int ds~[P(s)-P_\text{WD}(s)]}{\int ds~[P_\text{P}(s)-P_\text{WD}(s)]}
\label{eq:EtaObservable}
\end{equation}
based on the NND $P(s)$ where the integrals may be chosen \brc{including a proper argumentation} on
a suitable interval where one expects sensitivity of $P(s)$ on localization. According to our discussion
from \hyperref[sec:LocMeas]{Sect.\ \ref{sec:LocMeas}} one may take $s\in[0,s_{max}]$ with $s_{max}$
smaller than the smallest root of $P_\text{WD}(s)=P_\text{P}(s)$. But since our one-dimensional disordered system in use
does not converge to the Wigner surmise it is rather vague to apply \hyperref[eq:EtaObservable]{Eq.\ (\ref{eq:EtaObservable})}
in order to resolve the question on the impact of correlation on localization.

Nevertheless, the previous discussion invites us to introduce a similar measure for plotting the dependence
of localization on the parameter space $(\alpha,\kappa)$ which will be the last purpose
of this section on correlated disorder. We determine the \textit{deviation} of $P(s)$ from
$P_{\text{P}}$ as a quantity for the \textit{degree} of delocalization. Thus we define
\begin{equation}
L^2_\text{NND}(\kappa,\alpha)\equiv s^{-1}_\text{max}\int_0^{s_\text{max}}ds[P(s,\kappa,\alpha)-P_\text{P}(s)]^2
\label{eq:NNDDevFromPois}
\end{equation}
where $s_\text{max}$ introduces the numerically necessary restriction to a finite
range of $s$--values and we use $L_\text{NND}\equiv\sqrt{L^2_\text{NND}}$ as a localization
measure. Numerically, $L_\text{NND}$ features the striking advantage that it is sufficient to just compute
the eigenvalues of the Hamilton matrix $H_{s,ij}$ and therefore we are able to reach larger system sizes
$L$ with our code.

\begin{figure}[t!]
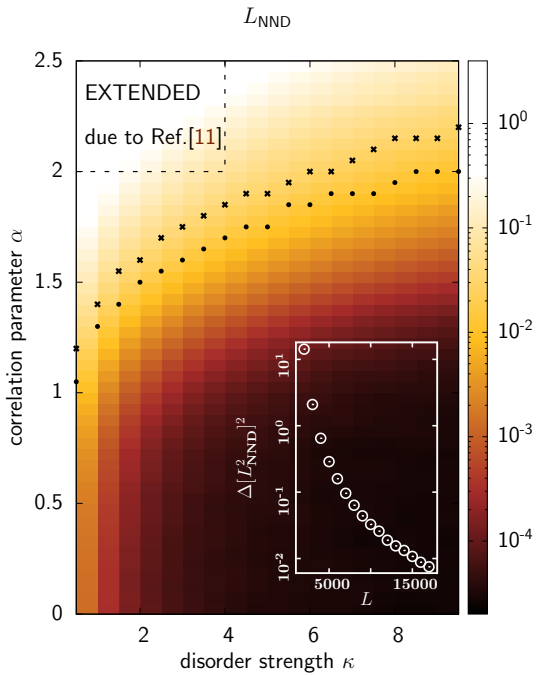

\centering
\resizebox{0.42\textwidth}{!}{\relsize{+3}\input FIG5}
\caption{\label{fig:NNDResKA}\brc{color online} $(\kappa,\alpha)$--dependence of
localization in the one-dimensional Anderson model from the perspective of a measure based on
the \textit{nearest neighbor spacing distribution} \brc{NND} of the Hamiltonian's
spectrum with $L=3\cdot10^{4}$. The finite size analysis with the measure
$\Delta[L^2_\text{NND}]^2$, defined in \hyperref[eq:DefErrLocMeas]{Eq.\ (\ref{eq:DefErrLocMeas})},
is shown as an inset \brc{white}. Furthermore we included the region which corresponds to
extended states according to the work of Shima et al.\cite{Shim04PhysRevB}\ who tackeled the
\textit{phase diagram} by means of states in the band center $E=0$ using the transfer matrix method.
The ($\bullet$) mark the minimal deviation of the NND from $P_\text{SP}$ \brc{cf.\ \hyperref[eq:SemPoisNND]{Eq.\ (\ref{eq:SemPoisNND})}}
for fixed disorder strength $\kappa$ and the ({\sf x}) indicate the same minima with respect to $P_\text{WD}$.}
\end{figure}

As we know from the previous considerations on the localization--delocalization crossover
for intermediate disorder \brc{$\kappa=1$, first part of \hyperref[sec:CorrDis_NumRes]{Sect.\ \ref{sec:CorrDis_NumRes}}},
correlation with $1\leq\alpha\leq2$ seems to be important for delocalization; varying $\alpha$ in $[0,2.5]$ will cover it. In the
limit $\kappa\to0$, i.e.\ vanishing disorder, it is obviously hardly possible to recover
Anderson's result for infinitely extended systems and arbitrary small disorder, since
we have to face the fact to be generally restricted to finite system sizes $L$ and
finite numerical precision. Therefore we investigate the domain $(\alpha,\kappa)\in[0,2.5]\times[0.5,10]$
which spans a wide range of disorder strengths $\kappa$ for the interesting amount
of correlation among the onsite disorder potential values $\epsilon_i$. 
The corresponding result is shown in \hyperref[fig:NNDResKA]{Fig.\ \ref{fig:NNDResKA}}
with $L=3\cdot10^4$
and the inset presents the finite size analysis we performed: Let $\Delta L$ denote the
\brc{fixed} difference between two system sizes $L$ and $L'$ such that $L=L'+\Delta L$
and we define
\begin{multline}
\Delta[L^2_{\text{NND}}]^2(L,\Delta L)\equiv \abs{\mathcal{A}}^{-1}\int_\mathcal{A} d\kappa d\alpha~\times\\
[L^2_{\text{NND}}(\kappa,\alpha,L)-L^2_\text{NND}(\kappa,\alpha,L')]^2
\label{eq:DefErrLocMeas}
\end{multline}
with $\abs{\mathcal{A}}=\abs{[0,2.5]\times[0.5,10]}=\Delta\kappa\Delta\alpha$ the area
of integration. The white inset plots that quantity versus $L$ having fixed $\Delta L=1000$; it
becomes clear that \hyperref[fig:NNDResKA]{Fig.\ \ref{fig:NNDResKA}} exhibits a \text{convergent} trend
for increasing $L$.

\begin{figure*}[t!]
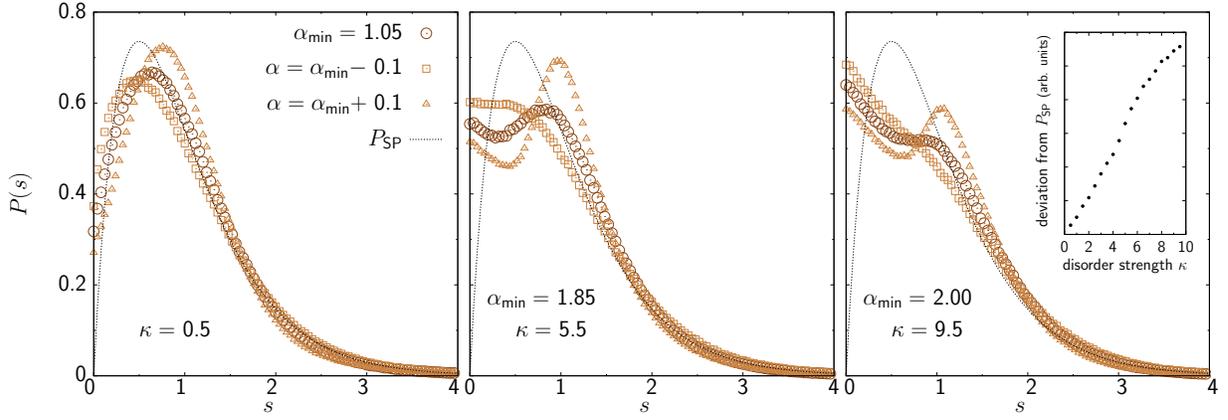

\centering
\resizebox{0.9\textwidth}{!}{\relsize{+5}\input FIG6}
\caption{\label{fig:NNDMinDevSPEx}\brc{color online} NNDs for different disorder strengths $\kappa$ \brc{$L=3\cdot10^4$}
to illustrate that the Semi-Poisson law $P_\text{SP}$ as a \textit{hybrid} between the Poisson distribution $P_\text{P}$ and
the \textit{Wigner surmise} $P_\text{WD}$ is not an appropriate critical distribution to indicate delocalization
for our model under consideration. The NND $P(s)$ with $\alpha_\text{min}$ \brc{$\odot$} is singled out by its \textit{minimal
deviation} from $P_\text{SP}$ for fixed $\kappa$ as described in the main text \brc{cf.\ also \brc{$\bullet$} in
\hyperref[fig:NNDResKA]{Fig.\ \ref{fig:NNDResKA}}}. Plotting $P(s)$ with $\alpha_\text{min}\pm0.1$ underlines
the two statements we emphasized: a) decreasing values of the NND for $s\to0$ \brc{\textit{level repulsion}} and b) development
of a peak of $P(s)$ around $s=1$ for increasing correlation parameter $\alpha$ due extended states \brc{increase of $P(1)$}.
The inset in the right panel shows the dependence of the minimal deviation from $P_\text{SP}$ with respect to the disorder strength.}
\end{figure*}

Before turning to a more detailed analysis of $L_\text{NND}(\kappa,\alpha)$ let us come back to
our comments on a critical distribution $P_c$. As stated before we do not expect the NND to converge
to $P_\text{WD}$ in the limit $\alpha\to\infty$ \brc{delocalized system}.
In our case it is not well justified to assume, e.g., the \textit{Semi-Poisson law}\cite{Jacq98PhysRevLett}
\begin{equation}
P_\text{SP}(s)\equiv4s\exp(-2s)
\label{eq:SemPoisNND}
\end{equation}
that resembles the linear increase of $P_\text{WD}(s)$ for $s\ll1$ and an exponential decay for
$s\gg1$ to be critical. Numerical evidence on our claim provides \hyperref[fig:NNDResKA]{Fig.\ \ref{fig:NNDResKA}}
where we plot both the minimal \textit{deviation} of the NND from $P_\text{SP}$ \brc{$\bullet$} and $P_\text{WD}$
\brc{\sf x} for fixed disorder strength $\kappa$. The notion \textit{deviation} is defined in total analogy to
\hyperref[eq:NNDDevFromPois]{Eq.\ (\ref{eq:NNDDevFromPois})}. In the case of the convergence of the NND
to $P_\text{WD}$ we would expect the minima \brc{\sf x} to appear at the maximal $\alpha$--value simulated ---
independent from the disorder strength $\kappa$. But, as visible in \hyperref[fig:NNDResKA]{Fig.\ \ref{fig:NNDResKA}},
the NND \textit{passes by} $P_\text{WD}$ below $\alpha\approx2.2$. But surprisingly, the deviation minima
of the NND from $P_\text{SP}$ \brc{$\bullet$} appear to lie in the crossover region $1<\alpha<2$
we emphasized during our previous numerical studies. Moreover, they follow the qualitative trend of our localization
measure $L_\text{NND}$ which one may naively exploit to argue in favor of $P_\text{SP}$ as a suitable critical distribution.
But a closer look on the precise shape of the NND with minimal deviation from $P_\text{SP}$ reveals significant differences
from the Semi-Poisson law that become even more obvious with increasing disorder strength $\kappa$. A corresponding
plot is shown in \hyperref[fig:NNDMinDevSPEx]{Fig.\ \ref{fig:NNDMinDevSPEx}}. Therefore, we do not support the
Semi-Poisson law $P_\text{SP}$ as a suitable critical distribution. In fact it is designed as a
\textit{hybrid} between the two limiting cases of a Poisson distribution $P_\text{P}$ and the Wigner surmise
$P_\text{WD}$ which does not apply to extended states in our specific model.

However, there has been a similar investigation to our \textit{phase diagram} in Ref.\ [\onlinecite{Shim04PhysRevB}]
and we would like to discuss some conclusions one may draw when comparing the two results now.
Shima et al.\ report numerical studies that support the existence of extended states for
$(\kappa,\alpha)\in[0,4]\times[2,5]$ \footnote{According to their notation $\kappa$ translates to $W$
and $\alpha$ to $p$.} on basis of a quantity $\Lambda$ similar to our LL evaluated at the band
center $e=0$ \brc{cf.\ their Fig.\ 6}. We separated the corresponding region of extended states in
\hyperref[fig:NNDResKA]{Fig.\ \ref{fig:NNDResKA}} by a dashed line. There are two main differences
we would like to point out with respect to our investigations of $L_\text{NND}(\kappa,\alpha)$:
First of all, the localization--delocalization crossover depends on the disorder strength
and second $L_\text{NND}$ varies over approximately two orders of magnitude in
$1\lesssim\alpha\lesssim2$ quite independently from the value $\kappa$. Therefore the evaluation
of the NND underlines the picture of a smooth crossover between localization and delocalization
drawn before. The qualitative result of delocalization for all $\kappa$--values under consideration is
in accordance with our line of reasoning that the disordered system has to eventually delocalize for
$\alpha\to\infty$ due to Bloch's theorem. Therefore it would be interesting to push the numerical
analysis of Shima et al.\ beyond their maximal correlation $\alpha\leftrightarrow p=5$. However, assuming that the deviation of the results
of Shima et al.\ and our investigations is not caused by some hidden technicality\footnote{While
we directly rely on \hyperref[eq:ProdCorrDis]{Eq.\ (\ref{eq:ProdCorrDis})} to produce correlated
disorder \brc{see also \hyperref[app:CorrDis]{App.\ \ref{app:CorrDis}}},
Ref.\ [\onlinecite{Shim04PhysRevB}] performs some sort of \textit{Fourier filtering method} to obtain
long-range correlated $\epsilon_i$-values.}, the discrepancy of our conclusions to Shima et al.\ may reveal
two useful \textit{lessons}:
\begin{enumerate}
\item It reminds us of the difficulty of a proper interpretation of the NND with respect to localization.
The quantities $\Lambda$ and LL, respectively, are much closer to the original picture
of localization drawn by P.\ W.\ Anderson and thus one may prefer it in cases where physical
intuition is hard to gain. The NND always needs reasonable understanding of the
system under examiniation.
\item While Shima et al.\ focused on the states in the band center of the disordered one-dimensional
system, the NND indirectly takes into account the \brc{averaged} localization property of all states.
Assuming that both quantities properly account for localization, it seems that, at least for
$(\kappa,\alpha)\in[4,8]\times[2,5]$, the properties of states with $E=0$ are not dominating
enough to show up when all states are included.
\end{enumerate}
As a closing remark we would like to note that the previous discussion points out that
the application of the NND as a localization measure is not as straightforward as the other quantities
in the particular system we are investigating. Even though its appealing property of
being basis independent and thus easily employable for studying interacting particles,
its interpretation remains complicated. Eventually, this observation prevents us from
utilizing it in order to detect localization of multiple particles and we would rather like
to argue for an intuitive observable in the section below.

\section{Localization in the Presence of Interaction}
\label{sec:LocInt}
\noindent In this second part of our discussion on delocalization in the one-dimensional Anderson
model we want to turn to the question how localization is effected by the presence
of interaction. This problem basically goes beyond the scope of the quantum physics
of $H_s$ originally encountered by Anderson, but --- as mentioned in the beginning ---
the question of the impact of interaction already attracted his attention in the late
1970s. Nevertheless the problem has remained a challenging topic and experiments
with BECs in optical lattices have re--raised the focus on it during the past few years.

With \hyperref[eq:AndHamMP]{Eq.\ (\ref{eq:AndHamMP})} at hand, namely
\begin{equation}
H_{mp}=\sum^L_{i=1}{\epsilon_i c^+_i c_i}+J\sum^{L-1}_{i=1}{c^+_i c_{i+1}}+\sum_{i,j=1}^{L}U_{ij}c^+_ic^+_jc_ic_j+h.c.~,
\label{eq:AndHamMP_rev}
\end{equation}
we introduce an interaction term $U_{ij}c^+_ic^+_jc_ic_j$ where for $U_{ij}=U_0\delta_{ij}$
one encounters the \textit{disordered Bose--Hubbard model} assuming bosonic
commutation relations for the lattice site annihilation \brc{creation} operators
$c_i^{(+)}$. $U_0=const.$ describes \textit{contact/onsite interaction}, i.e.\ particles interact
only when occupying the same site. But as we will carry out in \hyperref[sec:LocInt_ModInt]{Sect.\ \ref{sec:LocInt_ModInt}}
it is convenient to assume the more general case, \hyperref[eq:AndHamMP_rev]{Eq.\ (\ref{eq:AndHamMP_rev})},
to describe experiments with either Rydberg gases \cite{Li05PhysRevLett,Sing04PhysRevLett} or \brc{dipolar} BECs \cite{Grie05PhysRevLett}
where particles interact even if separated by a \brc{large} number of lattice sites. In particular, we use an algebraic
decaying interaction potential and therefore conceptually bridge from \textit{long-range}
correlation to \textit{long-range} interaction.

\subsection{Defining Localization}
\label{sec:LocInt_DefLoc}
\noindent Before proceeding we have to deal with the question on how to treat localization
for multiple \brc{interacting} particles, since localization was originally defined
for the single particle problem. The literature provides several approaches for a suitable
definition. E.g.\ one can use the \textit{Hausdorff} measure \cite{Aize09CommMathPhys}
as a \textit{distance} between two states $\psi_0$ and $\psi_t$ where the latter
is the former one evolved in time by the Schr\"odinger equation $i\hbar\partial_t\ket{\psi}=H_{mp}\ket{\psi}$.
In analogy to Anderson's initiating paper \cite{Ande58PhysRev} one can then ask, roughly
speaking, for the \textit{absence of diffusion} in terms of that distance. Another
idea is to relate localization to macroscopic observables and associate the notion
of localization to vanishing electrical conductivity \cite{Bask06AnnPhys}. Furthermore
one can directly study localization in Fock space as performed, e.g., in Ref.\ [\onlinecite{Berk98PhysRevLett}]. 
An ansatz independent of the specific form of the Hamiltonian --- and thus easily applicable
to the interacting problem --- is the analysis of spectral statistics \brc{cf.\ \hyperref[eq:NNDDef]{Eq.\ (\ref{eq:NNDDef})}} \cite{Cuev99PhysRevLett}
which is driven by the analogy between random matrices in quantum chaos and the random Hamilton
matrices due to disorder. But as extensively discussed in \hyperref[sec:CorrDis_NumRes]{Sect.\ \ref{sec:CorrDis_NumRes}}
its strength of being representation independent may come along with some difficulties due
to a proper interpretation of the corresponding results. Furthermore, the application of the NND
to systems of multi-particles shifts the argument of level repulsion given in \hyperref[sec:LocMeas]{Sect.\ \ref{sec:LocMeas}}
to localization in Fock space.

For our purpose we want to propose a rather intuitive solution inspired by a quantity
which seems to be natural under the scope of experiments with cold atoms and which
is connected to the particle density. Therefore
we try to motivate a projection of the many--particle state to the lattice sites
$i=1\dots L$ and apply a one--particle localization measure afterwards. Of course,
the procedure has to coincide with the notion of single particle localization
if we return to a set of Fock states with one particle at a given site.

Since one is able to directly image the density profile of Bose--Einstein condensates
in experiment \cite{Bill08nat} we consider the question on the probability $p_i$ to
find \textit{at least} one particle at site $i$. Using the language of second quantization
we define the \textit{projection operator}
\begin{equation}
\mathcal{P}_i\equiv\frac{c_i}{\sqrt{n_i+\delta_{0n_i}}}
\end{equation}
where $n_i$ is the eigenvalue of the \textit{number operator} $\hat{n}_i\equiv c_i^+c_i$
counting the number of bosons at site $i$ and the \textit{Kronecker delta} $\delta_{kl}$ ensures
that $\mathcal{P}_i$ is well defined for $n_i=0$. Given an eigenstate $\ket{E}$ of
$H_{mp}$ we can answer our question from above by
\begin{equation}
p_i(E)\propto\abs{\bra{E}(\mathcal{P}_i^+\mathcal{P}_i)\ket{E}}^2
\label{eq:DefProjToSiteBas}
\end{equation}
and the constant of proportionality is obtained by the normalization condition $1=\sum_{i=1}^Lp_i$.
Therefore we end up with a projection of the multi-particle state $\ket{E}$ onto the
site basis $\ket{i}$ by the \textit{probability distribution} $p_i(E)$. For the single particle problem this
projection is exactly given by the site basis representation of $\ket{E}$, i.e.\
$p_i(E)=\abs{\mv{i}{E}}^2$ whose properties we studied for correlated disorder above.
Hence the natural step to perform next is to apply the NSD or the IPR on $p_i$ 
\brc{$i=1\dots L$}. But there is an argument excluding the application of the NSD measure: Suppose, two
particles are localized at certain sites $i_1$ and $i_2$, i.e.\ $p_i$ is peaked for $i\in\{i_1,i_2\}$. If
$\abs{i_1-i_2}\sim L$ the NSD will yield values of order $1$ and we would refer to
the corresponding $\ket{E}$ as delocalized. Thus we exclusively consider the IPR
and our definition of localization turns, roughly speaking, to the association on
how many sites are occupied by the interacting particles.

\subsection{Modelling Interaction}
\label{sec:LocInt_ModInt}
\noindent As announced above we want to consider \textit{long-range} correlation inspired by exeriments
with dipolar gases. Since no \textit{distance} $\Delta x$ between neighboring sites enters the
theory we can think of the limit $L\to\infty$ twofold. It can either refer to approaching
an infinitely large system with finite $\Delta x$ or the continuum limit \brc{$\Delta x\to0$}
of a finite total system length $x=L\Delta x$ \brc{like in experiments}. The following
discussion prefers the second picture.

To properly model $U_{ij}$ let us consider
\begin{equation}
U_{ij}\equiv U(l)\sim l^{-3}\quad\text{for}\quad l\equiv\abs{i-j}\gg1
\end{equation}
according to the dipole--dipole coupling. However, for $l\sim1$ the interaction has
to be renormalized to reach a finite value at $l=0$ which corresponds to the \brc{Bose--Hubbard}
\textit{contact interaction strength} $U_0$. To respect these limits we use the $L$--dependent\footnote{We
choose the interaction $U_{ij}$ to dependent on the system size $L$ to ensure the
constance of the interaction within the picture of the continuum limit described
above when $L$ tends to infinity.} interaction
\begin{equation}
U_{ij}\equiv U^\pm_L(l)=\pm\left(\left[\frac{l}{\lambda_IL}\right]^3+\abs{U_0}^{-1}\right)^{-1},~
\text{\small$l=\abs{i-j}$}~.
\end{equation}
Note that $\lambda_I>0$ specifies the \textit{range} of interaction and the \textit{sign} $\pm$
determines repulsion and attraction, respectively.

\subsection{Two bosons with dipole--dipole coupling: Numerical results}
\label{sec:LocInt_TwoIntBos}
\noindent So far our considerations to define localization and the type of interaction
were rather general on the total particle number $n=\sum_{i=1}^Ln_i$ and the system
size $L$, respectively. But for numerical simulations we need to set up a practically
tractable situation: Since the dimension of the corresponding Fock space grows exponentially
with the total number of particles when assuming \textit{constant filling}\footnote{Constant
filling means $n\sim L$ and together with $\dim H_{mp}=(n+L-1)!/(n!(L-1)!)$ from elementary
combinatorics and Stirling's approximation $\log x!\sim x\log x-x$ we obtain
$\dim H_{mp}\sim a^L$ for $L\gg1$ and $a\sim1$.}, we investigate the specific case
of constant particle number $n$. Therefore the complexity class of $\dim H_{mp}$
shrinks to $\mathcal{O}(L^n)$ and in the specific case of two particles we obtain
\begin{equation}
\dim H_{mp}=\frac{L(L+1)}{2}=\mathcal{O}(L^2)~,\quad\text{$n=2$}\quad.
\end{equation}
Hence we are able to reach much larger system sizes $L$ in contrast to the situation
$L\sim n$ and we are especially interested in the \brc{well-established} toy model \cite{Shep94PhysRevLett,Imry95EurPhys,Frah95EurophysLett,Roem97PhysRevLett}
of two interacting particles since D. L. Shepelyansky and Y. Imry provide arguments
for the invariance of localization with respect to the sign of interaction which we
can check numerically. Of course, we do not expect that studying just
two interacting particles in a disordered potential will fully account for the effects
of finite densities in cold atom experiments, but before turning to \textit{full complexity}
we may gain intuition on the problem by means of this academic example. As we
will describe below there is some phenomenological reasoning that relates
delocalization by correlation and interaction.

As in \hyperref[sec:CorrDis_NumRes]{Sect.\ \ref{sec:CorrDis_NumRes}} we apply exact
numerical diagonalization to the Hamiltonian $H_{mp}$ and hence we have to specify
a suitable basis: We take $\{\ket{n_1\dots n_i\dots n_L}\}$ and \hyperref[app:MatrRepMP]{App.\ \ref{app:MatrRepMP}}
provides details on the explicit form of the Hamilton matrix for arbitrary size $L(L+1)/2$.
The inclusion of an interaction term $U_{ij}$ to the problem enlarges our space of
system parameters from $(\kappa,\alpha)$ to $(\kappa,\alpha,\pm u_0,\lambda_I)$, where
\begin{equation}
u_0\equiv\abs{U_0}/J
\end{equation}
accounts for the rescaling discussed in the beginning of \hyperref[sec:CorrDis_NumRes]{Sect.\ \ref{sec:CorrDis_NumRes}}.
We should concentrate on the new degrees of freedom $(\pm u_0,\lambda_I)$ and 
\begin{figure}[t!]
\centering
\includegraphics[width=.48\textwidth]{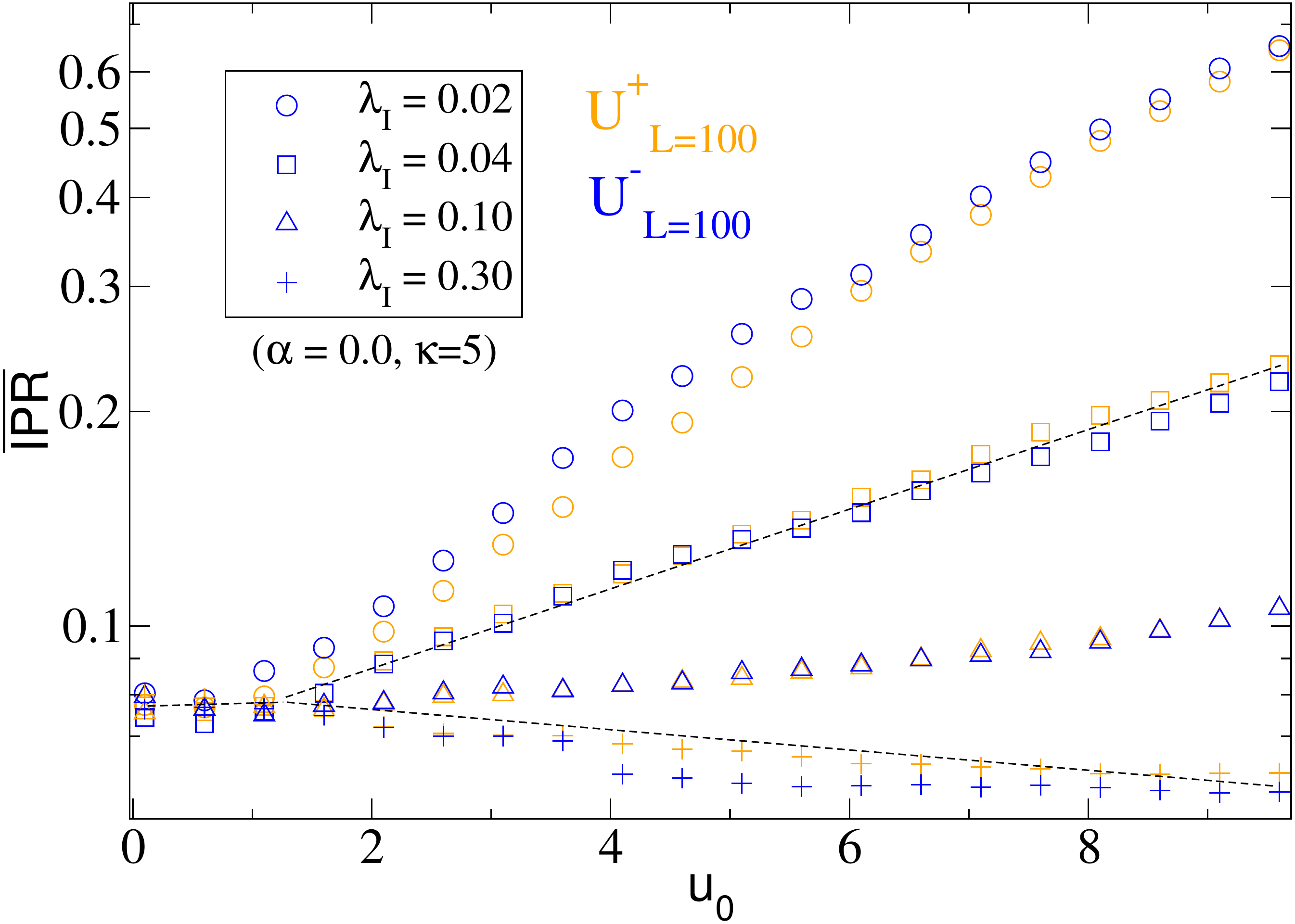}
\caption{\label{fig:IPRAvResUncorr}\brc{color online} DOS convoluted inverse participation
ratio $\overline{\text{IPR}}$ on the basis of the projection, \hyperref[eq:DefProjToSiteBas]{Eq.\ (\ref{eq:DefProjToSiteBas})}.
The quantity was evaluated for several \brc{rescaled} onsite interaction strengths
$u_0$ and interaction ranges $\lambda_I$ at fixed $\kappa=5$. It is remarkable that
the result seems to be independent of whether the dipole--dipole coupling is repulsive \brc{orange}
or attractive \brc{blue}. Furthermore the plot suggests that $\lambda_I$ is the crucial
parameter to trigger whether increasing interaction strength $u_0$ delocalizes or
localizes the system. Technically, we evaluated systems of size $L=100$ and averaged
over $30$ realizations. The dashed lines \brc{black} sketch the two branches we present
in \hyperref[fig:IPRAvResCorr]{Fig.\ \ref{fig:IPRAvResCorr}} when additionally considering correlation.}
\end{figure}
therefore we fix $\kappa$ as well as consider uncorrelated disorder $\alpha=0$ where Anderson localization
of all states is a proven fact in one dimension for vanishing interaction. To obtain
a measure that characterizes the localization property of the \textit{whole} system \brc{parameters $\kappa,\alpha$ fixed},
i.e.\ for all $\ket{E}$ satisfying $H_{mp}\ket{E}=E\ket{E}$ \brc{cf.\ \hyperref[fig:IPRRes]{Fig.\ \ref{fig:IPRRes}}}
we convolute $\text{IPR}(E)$ extracted from $p_i(E)$ with $\text{DOS}(E)$ as performed several times
during our numerical studies in \hyperref[sec:CorrDis_NumRes]{Sect.\ \ref{sec:CorrDis_NumRes}}
\brc{cf.\ \hyperref[eq:NSDAvDef]{Eq.\ (\ref{eq:NSDAvDef})}}: 
\begin{equation}
\overline{\text{IPR}}(\pm u_0,\lambda_I)\equiv\int dE~\text{IPR}(E)\text{DOS}(E)\quad.
\label{eq:IPRAvDef}
\end{equation} 
The corresponding result is shown in \hyperref[fig:IPRAvResUncorr]{Fig.\ \ref{fig:IPRAvResUncorr}}.
As independently predicted by Shepelyansky and Imry localization seems to be invariant under the
transformation $U_{ij}\to-U_{ij}$. Moreover, the range $\lambda_I$ determines whether
the increasing onsite interaction strength $u_0$ \brc{weakly} delocalizes or
tends to localize the two bosons. This effect sets in when $\abs{U_0}\sim J$, i.e.\ when the interaction
energy becomes comparable to the kinetic contributions in $H_{mp}$.

A plausible argument for the observation that our results yield independence from the sign of the mutual
particle interaction relies on a discrete \brc{spatial} symmetry of the Hubbard Hamiltonian \cite{Sakm10PhysRevA,Schn10arXiv}
which we apply to \hyperref[eq:AndHamMP_rev]{Eq.\ (\ref{eq:AndHamMP_rev})}. Suppose we transform the spatial
wave function $\mv{j}{E}$ such that all, say, odd site $j$ contributions are inverted and all even ones
are left unchanged. The corresponding operator in the basis $\{\ket{j}\}$ reads
\begin{equation}
\mathcal{U}\equiv\mathrm{diag}(\dots,-,+,-,+,-,+,\dots)=\mathcal{U}^+
\end{equation}
and satisfies $1=\mathcal{U}^2$, i.e.\ $\mathcal{U}$ is a unitary transformation. In terms of annihilation \brc{creation}
operators one obtains $\mathcal{U}c^{(+)}_j\mathcal{U}^+=(-)^jc^{(+)}_j=e^{i\pi j}c^{(+)}_j$. Now, the key observation is that
\begin{equation}
\mathcal{U}H_{mp}(U_0)\mathcal{U}^+=-H_{mp}(-U_0)
\label{eq:Symmtry}
\end{equation}
approximately holds when averaging over disorder realizations \brc{cf.\ \hyperref[eq:IPRAvDef]{Eq.\ (\ref{eq:IPRAvDef})}},
since on average for each set $\{\epsilon_j\}$ there will be another one $\{-\epsilon_j\}$ when $\alpha=0$.
More precisely, if we fix the disorder potential $\{\epsilon_j\}$ the solution $H_{mp}(-U_0)\ket{E'}=E'\ket{E'}$
is related to the solution of $H_{mp}(U_0)\ket{E}=E\ket{E}$ with $\{-\epsilon_j\}$ by the identifications
$E'=-E$ and $\ket{E'}=\mathcal{U}\ket{E}$ which obviously leaves physical observables/localization measures invariant.

Concerning the qualitative different localization properties of the interacting bosons with respect to the range $\lambda_I$
we would like to provide some phenomenological argument that may also draw a unified picture of delocalization by
\begin{figure}[t!]
\centering
\includegraphics[width=.48\textwidth]{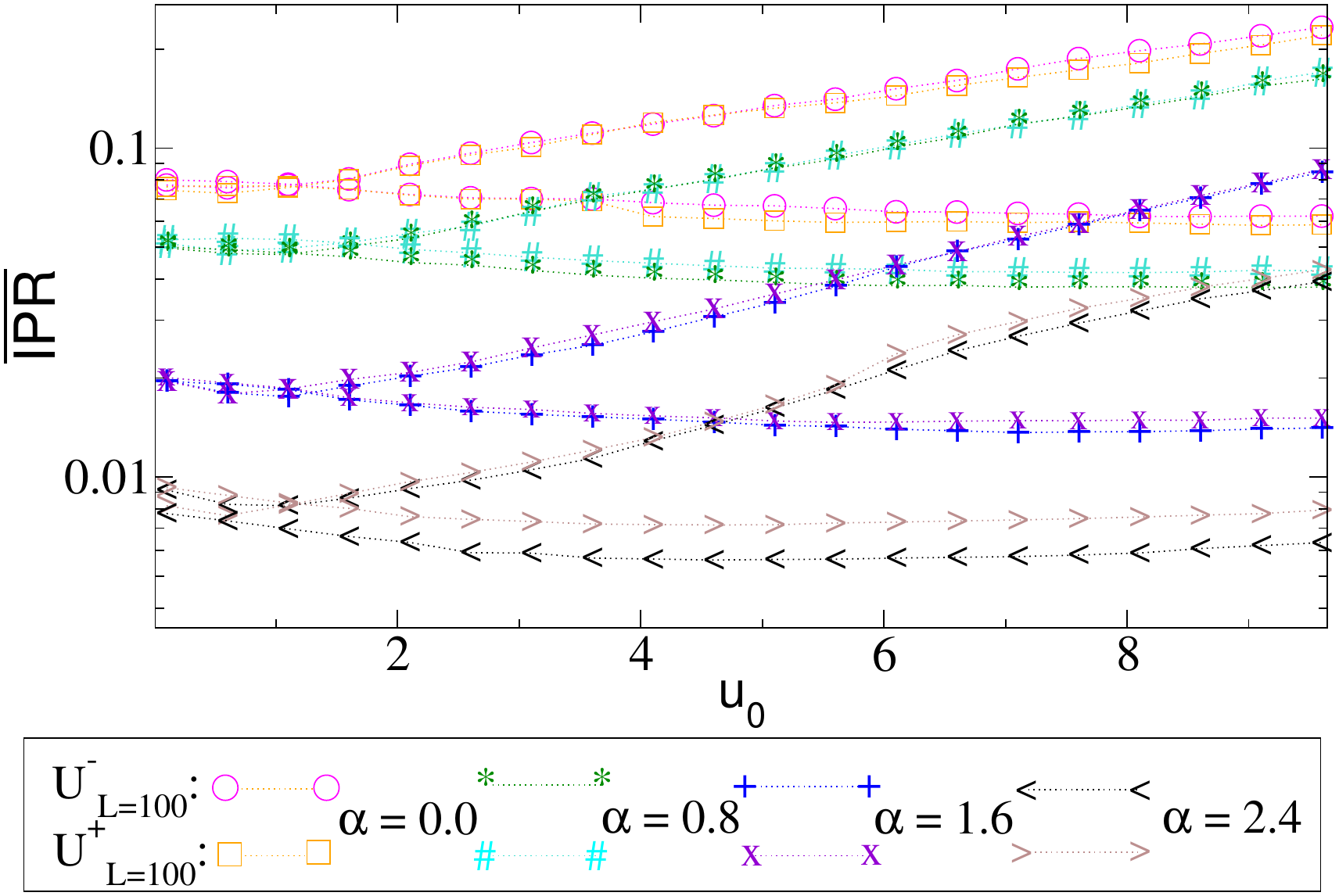}
\caption{\label{fig:IPRAvResCorr}\brc{color online} Impact of correlation \brc{$\alpha>0$,
$\alpha=0$ as a reference} on the localization property of two interacting bosons. We picked out two representative
branches \brc{$\lambda_I=0.04$ and $\lambda_I=0.3$} from \hyperref[fig:IPRAvResUncorr]{Fig.\ \ref{fig:IPRAvResUncorr}}
to demonstrate that, as expected, correlations tends to delocalize the system. Furthermore
we note that the qualtitative result is again independent of the sign of the interaction
and it seems that a feature from $\alpha=0$ stays untouched: Namely, small
interaction ranges $\lambda_I$ \brc{weakly} delocalize with increasing interaction
strength $u_0$ and with larger $\lambda_I$ states localize with respect to increasing
$u_0$. The numerical data were obtained at $\kappa=5$ from $30$ realizations with $L=100$
sites which we averaged over.}
\end{figure}
\begin{figure*}[t!]
\centering
\includegraphics[width=.8\textwidth]{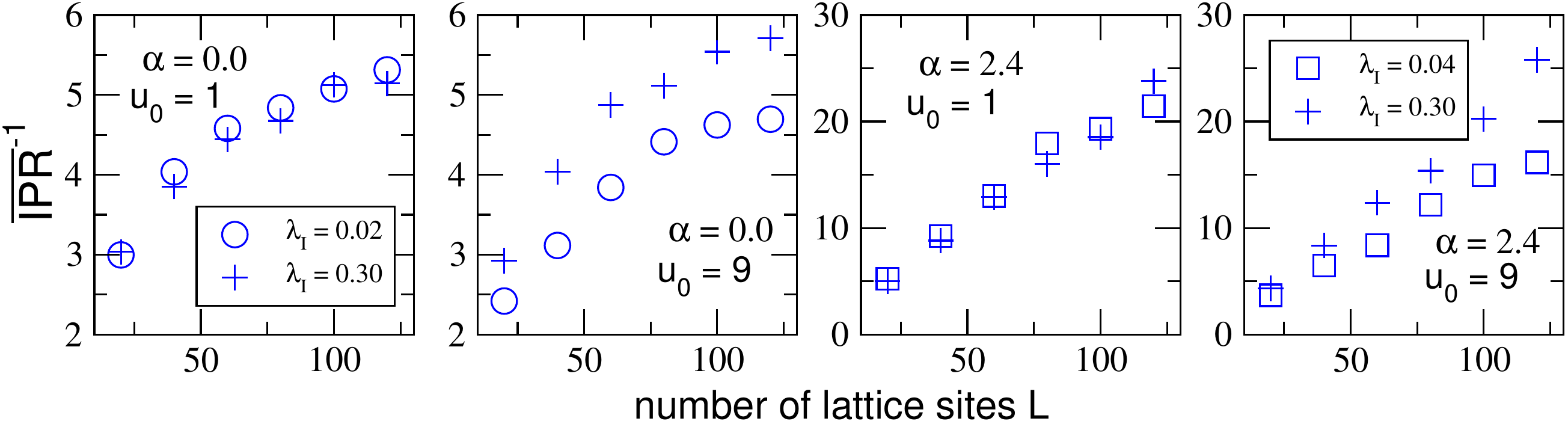}
\caption{\label{fig:FSSIPRResCorrDis}\brc{color online} Finite size analysis of the $\overline{\text{IPR}}^{-1}$
for representative model parameter values that specify the correlated disorder by $\alpha$ and the interaction by its
strength $u_0$ and range $\lambda_I$. Due to the observed sign symmetry we restrict to the case of attractive interaction;
the color encoding and point shape is kept as in \hyperref[fig:IPRAvResUncorr]{Fig.\ \ref{fig:IPRAvResUncorr}}. Instead
of plotting the averaged inverse participation ratio $\overline{\text{IPR}}$ itself, it is more appropriate to consider
its inverse since it directly relates to the number of occupied sites of the projected probability distribution $p_i$
\brc{cf.\ \hyperref[eq:IPR]{Eq.\ (\ref{eq:IPR})} and \hyperref[eq:DefProjToSiteBas]{Eq.\ (\ref{eq:DefProjToSiteBas})}}.}
\end{figure*}
correlation and interaction. Starting from the case of \textit{short-range}\footnote{We would like to note that there is a minor
inaccuracy in notion here. Although the algebraic decay of the interaction potetial $U_{ij}$ for $\abs{i-j}\sim L$
is always \textit{long-ranged} compared to e.g.\ an exponential decrease, in the following the terms \textit{short--}
and \textit{long-range} will refer to $\lambda_I\ll1$ and $\lambda_I\sim1$, respectively.} \brc{$\lambda_I\ll1$},
attractive interaction the enhancement of localization may result from the effect of \textit{lumping}: Assuming sufficiently
strong interaction compared to the kinetic energy \brc{$U_0\gtrsim J$}, once the particles are nearby, they will be
tightly bound in space and the quantum dynamics should be governed by this \textit{localized} behavior. Due to the
\textit{approximative} invariance of the problem to the interaction's sign the situation \brc{surprisingly} stays
unchanged even for repulsive interacting bosons. But according to \hyperref[fig:IPRAvResUncorr]{Fig.\ \ref{fig:IPRAvResUncorr}}
the effect of long-range interaction \brc{$\lambda_I=0.3$} is qualitatively opposite to the short-range case
and we numerically checked that the trend to delocalize with increasing interaction stays up to $\lambda_I=1$
\brc{not shown here}. One may understand this behavior on the basis of the Fock space Hamilton matrix
representation $H_{mp,ff'}$ \brc{cf.\ \hyperref[app:MatrRepMP]{App.\ \ref{app:MatrRepMP}},
\hyperref[eq:DefMultHamMat]{Eq.\ (\ref{eq:DefMultHamMat})} and \hyperref[tab:MatrStruc]{Tab.\ \ref{tab:MatrStruc}}}.
Since disorder $\epsilon_i$ and interaction $U_{ij}$ both contribute to the diagonal elements $H_{mp,ff}$ we may effectively
relate the resulting multi-particle Hamiltonian matrix to a corresponding one-dimensional, non-interacting
system of size $L(L+1)/2$ with modified disorder $\tilde{\epsilon}_{i'=1\dots L(L+1)/2}$ and extended dynamics beyond the 
nearest neighbor hopping\footnote{We would like to remind that Anderson's initiating paper\cite{Ande58PhysRev} did
not restrict to the case of nearest neighbor hopping. That was also the reason why we mentioned to deal with
a certain \textit{subclass} of Hamiltonians, \hyperref[eq:AndHamSP]{Eq.\ (\ref{eq:AndHamSP})}, in
\hyperref[sec:IntroMot]{Sect.\ \ref{sec:IntroMot}}.}. While short-range interaction just contributes
to a few diagonal entries of $H_{mp,ff'}$, an increasing interaction range $\lambda_I$ affects more and
more of those matrix elements. Due to the deterministic character of the $U_{ij}$ \brc{smooth, algebraic
long-range decay} it will effectively correlate the $\tilde{\epsilon}_{i'}$ of the non-interacting analoguous
which we know to yield delocalization for increasing long-range correlation.

Finally, we want to include long--range correlation \brc{$\alpha>0$} into our toy model of two interacting bosons
to directly render the impact of correlation within our localization framework. In \hyperref[fig:IPRAvResCorr]{Fig.\ \ref{fig:IPRAvResCorr}}
we plot the uncorrelated case from \hyperref[fig:IPRAvResUncorr]{Fig.\ \ref{fig:IPRAvResUncorr}}
for two different interaction ranges $\lambda_I$ as a reference and, therefore, it becomes clear
that correlation among the onsite disorder yields delocalization --- as we would expect
from our previous experience so far \brc{cf.\ \hyperref[sec:CorrDis_NumRes]{Sect.\ \ref{sec:CorrDis_NumRes}}
as well as our phenomenological line of reasoning from the preceeding paragraph}. However, the feature that \textit{short--range}
\brc{$\lambda_I=0.04$} interaction localizes and \textit{long-range} \brc{$\lambda_I=0.3$} interaction
seems to delocalize the two bosons for increasing \brc{rescaled} interaction strength $u_0$ is similar to the
uncorrelated case.

Again, the result seems to be independent of the sign of interaction. Since we have introduced
correlation among the $\epsilon_i$ our argument according to
\hyperref[eq:Symmtry]{Eq.\ (\ref{eq:Symmtry})} should not be valid in general.
Therefore we suppose the specific disorder model to intrinsically fulfill the necessary
assumption from above. Referring to the limiting case $\alpha\to\infty$ we recognize that $\epsilon_i$
becomes $\cos$-like where $\epsilon_i\to-\epsilon_i$ just shifts the onsite potential by a phase value $\pi$.

We would like to close our discussion by adding comments on the finite size analysis done as well as
comparing our investigations to similar literature, especially Ref.\ [\onlinecite{Duke09NJP}], who consider
spinless fermions that interact when nearest neighbors to each other --- an analogous to the \brc{short-range}
onsite interaction scenario for bosons. Dukesz et al.\ employ the same disorder potential as us and study, among others,
the interplay between long-range correlated disorder and interaction for two particles and refer to it as the
\textit{dilute limit}. Their localization detection measure $\ev{\text{NPC}}$ is essentially the inverse
of our $\overline{\text{IPR}}$, but with the decisive distinction of being applied directly in Fock space ---
without any projection back to the lattice we advertised in \hyperref[sec:LocInt_DefLoc]{Sect.\ \ref{sec:LocInt_DefLoc}}.
Therefore we will be restricted to a comparison of qualitative features.

Concerning the left upper panel of Fig.\ 6 in Dukesz et al.\ we notice \brc{apart from minor deviation for small
interaction strength denoted by $\Delta$} the trend of enhanced localization \brc{decreasing $\ev{\text{NPC}}$}
for increasing interaction strength. A feature that is confirmed by \hyperref[fig:IPRAvResUncorr]{Fig.\ \ref{fig:IPRAvResUncorr}}
for sufficiently small interaction range $\lambda_I$ \brc{increasing $\overline{\text{IPR}}$}. 
Aside the non-interacting case \brc{$u_0\leftrightarrow\Delta=0$} this qualtitative trend remains when long-range
correlation is involved: According to fixed $\alpha$ the different $\ev{\text{NPC}}$--curves in
the lower left panel of Fig.\ 6 decrease in magnitude. Moreover, each single curve supports enhanced
delocalization for increasing correlation parameter $0\lesssim\alpha\lesssim4$.

To study effects of finite size it is actually wise to investigate $\overline{\text{IPR}}^{-1}$, since
our discussion on \hyperref[eq:IPR]{Eq.\ (\ref{eq:IPR})} suggests that it provides some notion on
the effective number of occupied sites. It is perhaps that stage where one benefits from the proposed
projection prescription of \hyperref[sec:LocInt_DefLoc]{Sect.\ \ref{sec:LocInt_DefLoc}} again.
While $\ev{\text{NPC}}$ measures \textit{occupation} of the eigenstates $\ket{E}$ in Fock space
our $\overline{\text{IPR}}^{-1}$ directly accounts for the projected probability distribution
$p_i$ \brc{cf.\ \hyperref[eq:DefProjToSiteBas]{Eq.\ (\ref{eq:DefProjToSiteBas})}} on the one-dimensional
lattice with $L$ sites. However, in order to establish a finite size analysis we depicted representative
interaction and correlation parameter values $(\alpha,\kappa,u_0,\lambda_I)$ and studied the
$\overline{\text{IPR}}$ with increasing system size $L$ in \hyperref[fig:FSSIPRResCorrDis]{Fig.\ \ref{fig:FSSIPRResCorrDis}}.
For uncorrelated \brc{$\alpha=0.0$} and correlated \brc{$\alpha=2.4$} disorder we chose two different
interaction strengths where there is no significant difference in the $\overline{\text{IPR}}$
with respect to \hyperref[fig:IPRAvResCorr]{Fig.\ \ref{fig:IPRAvResCorr}}: a) $u_0=1$ and
b) where the interaction range $\lambda_I$ significantly splits the $\overline{\text{IPR}}$ \brc{$u_0=9$},
respectively.

In the case of uncorrelated disorder \brc{left and middle left panel of \hyperref[fig:FSSIPRResCorrDis]{Fig.\ \ref{fig:FSSIPRResCorrDis}}}
the averaged number of occupied sites $\tilde{L}=\overline{\text{IPR}}^{-1}$ increases less than
linear with linear increasing system size. Thus the extrapolation to the thermodynamic limit
suggests localized states as for the non-interacting system\footnote{This conclusion is opposite to
Dukesz et al.\ who present their results in the lower right panel of Fig.\ 7 in Ref.\ [\onlinecite{Duke09NJP}].
It would be interesting to extend there study to larger system sizes beyond $L=60$. On the other hand
the deviation of the results suggest that a proper projection to the lattice is perhaps crucial
in order to investigate localization of multiple particles.}. However, we would like to point
out the increasing deviation of $\tilde{L}$ between short-range and long-range interaction for
strong interaction strength $u_0$ which we would like to phrase as \textit{enhanced delocalization
by long-range interaction} and which we suggest to keep in mind for further investigations of interacting
particles in disordered media. If we turn to the \textit{strongly correlated regime} \brc{$\alpha>2$} we
definitely observe a qualitative difference in the finite size analysis of $\tilde{L}$. Now the $\overline{\text{IPR}}^{-1}$
seems to grow linearly in $L$ and thus the relative number of occupied sites $\tilde{L}/L$ stays constant
for $L\to\infty$ assuming that the observed trend remains for $L>120$. Hence strong correlation 
among the disorder potential $\epsilon_i$ delocalizes in analogy to the numerical experience from the non-interacting
analysis, \hyperref[sec:CorrDis_NumRes]{Sect.\ \ref{sec:CorrDis_NumRes}}. But again, in the
case of $u_0=9$ the curves $\tilde{L}(L)$ increasingly deviate for the scenario of short-- and long-range
interaction, which verifies our proposal of enhanced delocalization by long-range interaction
and we scetched a potential \brc{phenomenological} reason by means of a \brc{rough} correspondence
to a non-interacting disordered system above.

In summary, we hope that our data demonstrated that there is a \textit{complex interplay} between correlation
and interaction which by means of the Hamilton matrix structure may arise from a similar origin, namely
the correlation of the diagonal elements. Furthermore, both effects may compensate each other as seen
for $\left.\overline{\text{IPR}}\right|_{\alpha=0,\lambda_I=0.04}$
and $\left.\overline{\text{IPR}}\right|_{\alpha=1.6,\lambda_I=0.3}$ at $u_0\approx7.5$
in \hyperref[fig:IPRAvResCorr]{Fig.\ \ref{fig:IPRAvResCorr}}. The numerical study
of the interacting bosons supports the independence of localization from the sign
of interaction within our picture of multi-particle localization, and we hope
that the rather academic treatment of two interacting bosons reveals
in experiments with cold gases\footnote{For
this purpose it is obviously necessary to produce correlated long-range disorder
with algebraic power spectrum which could be a challenging task for experiments.
Via \textit{Feshbach resonances} \cite{Kett08Proc} it is possible to tune the interaction strength
$u_0$, but the variation of the range $\lambda_I$ is perhaps difficult to establish.}; at least
in principle in the limit of sufficiently low densities. Neverthless we want to remind that all ideas are
based on rather phenomenological reasoning and numerical simulations where we are able to point
out qualitative trends only. Therefore our results can just pave the way for a more profound
understanding on theoretical grounds.

\section{Conclusion \& Perspectives}
\noindent To conclude our investigations we briefly summarize what was achieved within our study
of the one-dimensional Anderson model. In the first place we analyzed two mechanisms
that yield delocalization, namely a correlated disorder potential and
mutual interaction between two bosons. By reviewing a known model of correlated disorder from the perspective
of exact numerical diagonalization and different \brc{well-established} localization
measures, we showed that the process of delocalization is much more complex than observed
up to now: Namely, there are reasonable arguments to consider $\alpha=1$ as important for the
process of delocalization by correlation. Nevertheless the general trend that increasing correlation yields
delocalization was confirmed and localized states approximately vanish for $\alpha\gtrsim2$. In addition
we were able to numerically establish the full parameter dependence of the model system by means of the
nearest neighbor distribution \brc{NND} and therefore we went beyond the analysis that focuses on states
in the band center. The results suggest that the system eventually becomes delocalized for sufficient large
correlation, independent from the disorder strength $\kappa$ which is in accordance with Bloch's theorem.
Nevertheless, we extensively discussed the usage of the NND and tried to convince that it is far less obvious
to utilize it as an localization detection measure.

Therefore we introduced a general idea to define localization for multi-particle states of bosons by means of the
phenomenology of experiments with cold quantum gases. The examination of two long-range interacting bosons
\begin{figure*}[t!]
\centering
\includegraphics[width=.28\textwidth]{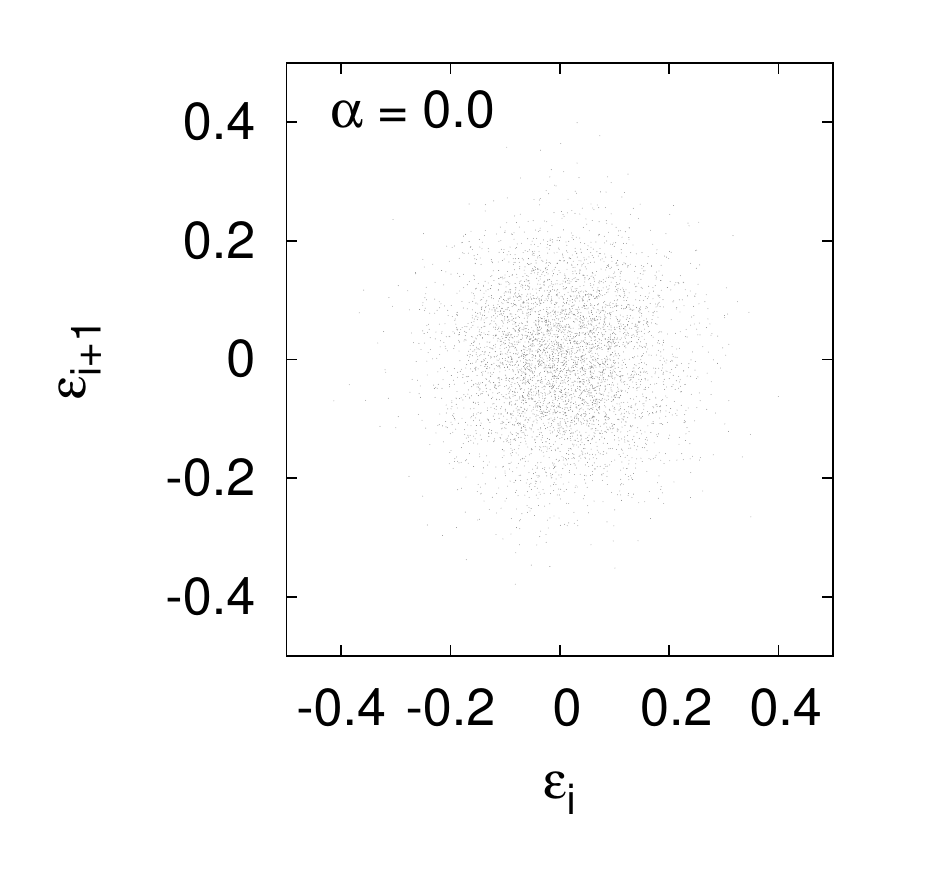}
\includegraphics[width=.225\textwidth]{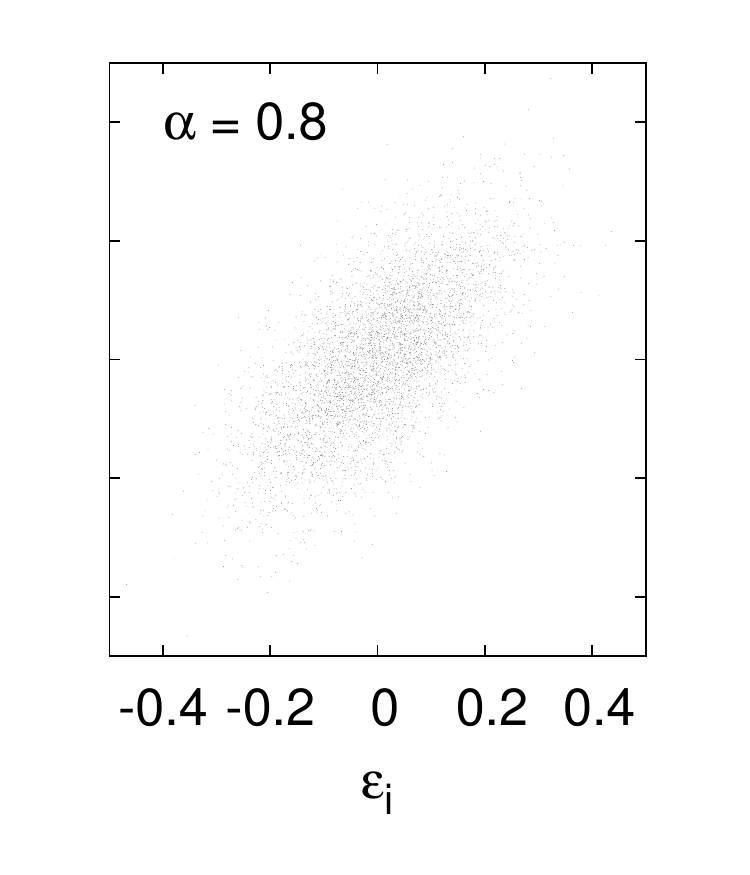}
\includegraphics[width=.225\textwidth]{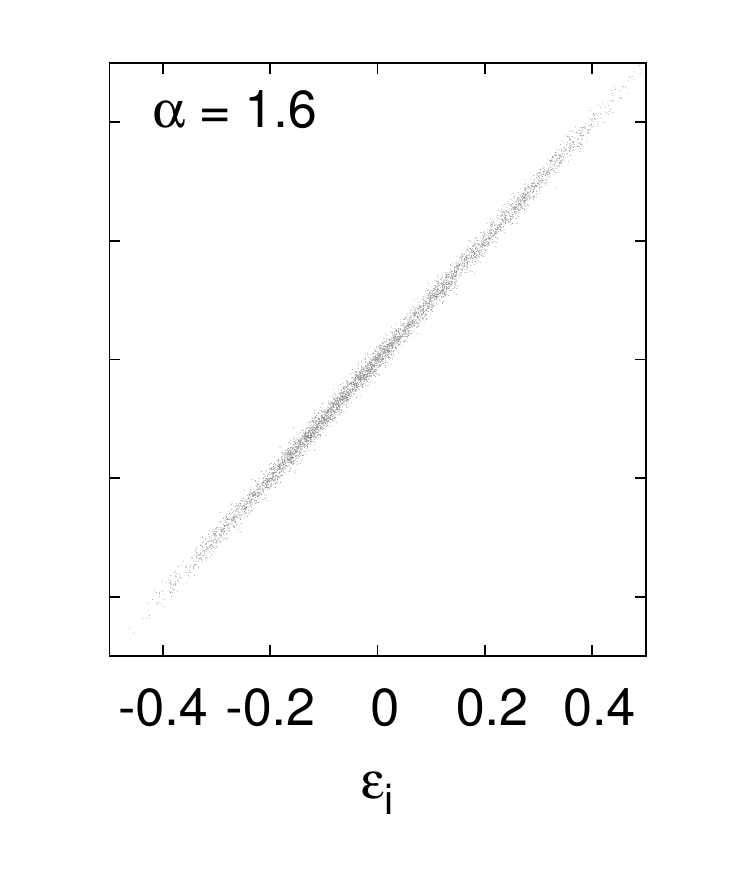}
\caption{\label{fig:SpecTestCorrDis}Illustration of \textit{autocorrelation} \brc{cf.\ \hyperref[eq:DefAutoCorrFkt]{Eq.\ (\ref{eq:DefAutoCorrFkt})}}
among the onsite disorder $\epsilon_i$, \hyperref[eq:ProdCorrDis]{Eq.\ (\ref{eq:ProdCorrDis})}. We
evaluated $5000$ points $(\epsilon_i,\epsilon_{i+1})$ with $N=10^5$ \brc{cf.\ \hyperref[eq:ProdCorrDis]{Eq.\ (\ref{eq:ProdCorrDis})}}.
An increasing correlation parameter $\alpha$ \textit{deforms} the \textit{rotationally invariant}
density of points $(\epsilon_i,\epsilon_{i+1})$ to a linear dependence between $\epsilon_i$
and $\epsilon_{i+1}$ which can be quantitatively detected by the \brc{linear} autocorrelation
$C(d)$. Rotational invariance yields $C(1)=0$ and linear dependence between $\epsilon_i$
and $\epsilon_{i+1}$ implies $C(1)=\pm1$. Note, that even in the case $\alpha=0$
our model disorder is \underline{not} totally homogeneous in $[-0.5,0.5]\times[-0.5,0.5]$.}
\end{figure*}
on a finite, one-dimensional lattice with \brc{perfect} disorder confirmed the conjecture that the localization property does
not dependent on the sign of interaction. We showed that this phenomenon leads back to a discrete symmetry arising
when averaging over disorder realizations. Finally the \textit{complex interplay} of correlation and interaction was studied
numerically and by means of phenomenological reasoning on the basis of the Hamilton matrix structure we argued
how to relate the impact of interaction to delocalization by correlation known from the non-interacting system.

Although our contribution provides some new insights to the phenomenon of localization, unsolved aspects remain.
More precisely, a solid theoretical description of the crossover from localization to delocalization when tuning the
\brc{nearly} perfect disorder to the \textit{Bloch--like} situation of a highly correlated potential
is desired. Moreover a more profound understanding of the numerically observed effects of interaction
on localization is imperative to obtain further insight to delocalization/localization in the presence of many interacting
particles.

\begin{acknowledgements}
\noindent We want to thank Tobias Paul for his collaboration on issues concerning the correlated
disorder part of this paper. We are grateful for computational resources provided by the
\textsl{bwGRiD} of the federal state Baden-W\"urttemberg \brc{Germany} and for support by
the \textsl{Heidelberg Center for Quantum Dynamics} as well as the \textsl{Heidelberg Graduate
School of Fundamental Physics} \brc{grant number GSC 129/1}.
\end{acknowledgements}

\appendix
\section{Numerical Details}
\noindent This appendix is dedicated to computational details of our work. We especially
address the correlated disorder used in the main part and an explicit calculation
of the Hamiltonian matrix of two interacting bosons for an arbitrary \brc{finite} number
of lattice sites.

\subsection{Correlated disorder}
\label{app:CorrDis}
\noindent Here, we provide some properties of the correlated disorder potential $\epsilon_i$
and as a general remark we mention that all our numerical results were averaged
over a number of disorder realizations. If we exactly follow the prescription
\hyperref[eq:ProdCorrDis]{Eq.\ (\ref{eq:ProdCorrDis})} to obtain $L$ values $\epsilon_i$
with correlation parameter $\alpha$ the quantity
\begin{equation}
\delta\equiv\max_i\epsilon_i-\min_i\epsilon_i\quad,\quad i=1\dots L
\end{equation}
will not be constant in general. To ensure the same disorder strength
$\kappa$ for all disorder realizations and system sizes, we always normalize a given chain of
$\epsilon_i$ values by
\begin{equation}
\epsilon_i\to\epsilon_i-\min_i\epsilon_i\to\frac{\Delta}{\delta}\epsilon_i\to\epsilon_i-\frac{\Delta}{2}
\end{equation}
to have $\epsilon_i\in[-\frac{\Delta}{2},\frac{\Delta}{2}]$. See also Refs. [\onlinecite{Kant00CommentOnMouraLyra},\onlinecite{Mour00ReplyOnMouraLyra}]
for a related discussion.

Since we mentioned that $\alpha=0$ refers to \textit{nearly} perfect disorder, we
provide an intuition on that statement in \hyperref[fig:SpecTestCorrDis]{Fig.\ \ref{fig:SpecTestCorrDis}}
where we illustrate increasing correlation by plotting $\epsilon_i$ versus $\epsilon_{i+1}$
for different \brc{increasing} $\alpha$--values: For $\alpha=0$ the \textit{point density}
is approximately \textit{rotationally invariant} around $(\epsilon_i,\epsilon_{i+1})=(0,0)$
and for the \textit{autocorrelation} function
\begin{equation}
C(d)\equiv\lim_{L\to\infty}
\frac{\ev{\epsilon_i\epsilon_{i+d}}_L-\ev{\epsilon_i}_L\ev{\epsilon_{i+d}}_L}
{\sigma_{\epsilon_i,L}\sigma_{\epsilon_{i+d},L}}\quad,
\label{eq:DefAutoCorrFkt}
\end{equation}
where
\begin{equation}
\ev{\epsilon_i}_L\equiv\frac{1}{L}\sum_{i=1}^L\epsilon_i\quad\text{and}\quad
\sigma^2_{\epsilon_i,L}\equiv\ev{\epsilon^2_i}_L-\ev{\epsilon_i}^2_L\quad,
\end{equation}
we find\footnote{In general, correlation between $\epsilon_i$ and $\epsilon_{i+d}$
is given by $C(d)$.} $C(1)\approx0$ for $\alpha=0$. By means of \hyperref[eq:DefAutoCorrFkt]{Eq.\ (\ref{eq:DefAutoCorrFkt})}
we can state that the disorder is \textit{uncorrelated}. But as obvious from \hyperref[fig:SpecTestCorrDis]{Fig.\ \ref{fig:SpecTestCorrDis}}
the whole plane $(\epsilon_i,\epsilon_{i+1})\in[-0.5,0.5]\times[-0.5,0.5]$ \brc{$\Delta=1$}
is not uniformly filled with points $(\epsilon_i,\epsilon_{i+1})$ and therefore an
arbitrary onsite potential value $\epsilon_i$ can not obviously be followed by
any possible $\epsilon_{i+1}$. Thus we refer to the case $\alpha=0$ as \textit{weakly}
correlated. On increasing correlation we observe that the $(\epsilon_i,\epsilon_{i+1})$
become more and more \textit{linearly correlated}, i.e.\ $\epsilon_i\approx const.\cdot\epsilon_{i+1}$
and thus $C(1)\approx1$.

Let us finally sketch one feature of the $\epsilon_i$ that arises when approximately
computing $C(d)$ to demonstrate the \textit{long-range} character of the correlated
disorder. Let
\begin{equation}
\epsilon(x_i)=\int_0^\infty dk~k^{-\alpha/2}\cos[kx_i+\phi_k],\quad x_i=i\Delta x
\end{equation}
be the continuous version of \hyperref[eq:ProdCorrDis]{Eq.\ (\ref{eq:ProdCorrDis})}
and
\begin{equation}
C(d)\propto\lim_{X\to\infty}\frac{1}{X}\int_0^Xdx~\epsilon(x)\epsilon(x+d)
\end{equation}
that of \hyperref[eq:DefAutoCorrFkt]{Eq.\ (\ref{eq:DefAutoCorrFkt})}. Hence we obtain
\begin{equation}
C(d)\propto\frac{1}{2}\int_0^\infty dk~k^{-\alpha}\cos(kd)
\end{equation}
which can be analytically solved for $\alpha\in(0,1)$ and we end up with
\begin{equation}
C(d)\propto d^{\alpha-1}\quad0<\alpha<1\quad.
\end{equation}
Hence we have a direct argument at hand why the $\epsilon_i$ are \textit{long-range}
correlated: Its spatial autocorrelation decays algebraically \brc{at least in the
continuum limit for $\alpha\in(0,1)$} such that increasing $\alpha$ yields a slower decay with distance
$d$. Within this picture $\alpha=1$ is also \textit{highly} correlated and one may argue
that delocalization occurs below $\alpha=2$ which is qualitatively supported by our
localization measures although the LL agrees pretty well with other results from
the literature \brc{cf.\ discussion in \hyperref[sec:CorrDis_NumRes]{Sect.\ \ref{sec:CorrDis_NumRes}}}.

\subsection{Matrix representation of two interacting bosons in the modified/interacting Anderson model}
\label{app:MatrRepMP}
\noindent For the numerical results on two interacting particles we developed a rather general
scheme of the matrix representation in Fock space which allows us to set the numerics
for an arbitrary number of sites. Moreover the matrix is banded with a width $\sim L$
which helps to increase the efficiency of the diagonalization routine. As we were
inspired by an experimental setup, we used \textit{hard--wall/open} boundary conditions,
i.e.\ there is no \brc{kinetic} hopping element $J$ \brc{cf.\ beginning of \hyperref[sec:IntroMot]{Sect.\ (\ref{sec:IntroMot})}}
from site $i=1$ to site $L$ and vice versa.

We obtained the matrix structure presented in \hyperref[tab:MatrStruc]{Tab.\ \ref{tab:MatrStruc}}
\begin{table*}
\centering
\begin{equation}
\footnotesize
\left[\begin{array}{ccccccccccccccccccccc}
x&\times&\multicolumn{7}{l}{\xleftarrow{\hspace*{4cm}}}&\multicolumn{4}{c}{\mathcal{O}(L^2)}&\multicolumn{8}{r}{\xrightarrow{\hspace*{4cm}}}\\
\times&x&\times&&&&\times&&&&&&&&&&\\
&\times&x&\times&&&\uparrow&\times&&&&&&&&&&&&&\\
&&\times&x&\times&&&&\times&&&&&&&&&&&&\\
&&&\times&x&\times&{\scriptstyle\mathcal{O}(L)}&&&\times&&&&&&&&&&&\\
&&&&\times&x&\downarrow&&&&\times&&&&&&&&&&\\
&\times&&&&&x&\times&&&&&&&&&&&&&\\
&&\times&&&&\times&x&\times&&&\times&&&&&&&&&\\
&&&\times&&&&\times&x&\times&&&\times&&&&&&&&\\
&&&&\times&&&&\times&x&\times&&&\times&&&&&&&\\
&&&&&\times&&&&\times&x&&&&\times&&&&&&\\
&&&&&&&\times&&&&x&\times&&&&&&&&\\
&&&&&&&&\times&&&\times&x&\times&&\times&&&&&\\
&&&&&&&&&\times&&&\times&x&\times&&\times&&&&\\
&&&&&&&&&&\times&&&\times&x&&&\times&&&\\
&&&&&&&&&&&&\times&&&x&\times&&&&\\
&&&&&&&&&&&&&\times&&\times&x&\times&\times&&\\
&&&&&&&&&&&&&&\times&&\times&x&&\times&\\
&&&&&&&&&&&&&&&&\times&&x&\times&\\
&&&&&&&&&&&&&&&&&\times&\times&x&\times\\
&&&&&&&&&&&&&&&&&&&\times&x\\
\end{array}\right]
\normalsize
\end{equation}
\caption{\label{tab:MatrStruc} Matrix structure of the multi-particle
Hamiltonian $H_{mp}$ in the 2-particle basis, \hyperref[eq:2PartBas]{Eq.\ (\ref{eq:2PartBas})} with $L=6$ sites.}
\end{table*}
where we labeled and ordered the Fock states according to
\begin{eqnarray}
\ket{1}_F&=&\ket{200\dots0}\nonumber\\
\ket{2}_F&=&\ket{110\dots0}\nonumber\\
\ket{3}_F&=&\ket{101\dots0}\nonumber\\
&\vdots&\nonumber\\
\ket{L}_F&=&\ket{100\dots1}\nonumber\\
\ket{L+1}_F&=&\ket{020\dots0}\nonumber\\
&\vdots&\nonumber\\
\ket{L(L+1)/2}_F&=&\ket{000\dots2}
\label{eq:2PartBas}
\end{eqnarray}
with $\ket{n_1n_2\dots n_L}$ indicating the number $n_i$ of bosons at site $i$ and
the matrix structure was exemplarily evaluated for $L=6$. Entries $\times$ denote
\brc{constant} kinetic off--diagonal values and $x$ refers to the diagonal elements
that are determined by the correlated disorder terms $\epsilon_i$ and the interaction
potential $U_{ij}$. If we write the Hamiltonian matrix elemtens as
\begin{equation}
H_{mp,ff'}\equiv\ipop{f}{H_{mp}}{f'}_F
\label{eq:DefMultHamMat}
\end{equation}
the generalization to an arbitrary number of sites $L$ reads:
\begin{widetext}
\begin{center}
\begin{tabular}{l*{7}{p{0.085\textwidth}}p{0.105\textwidth}p{0.085\textwidth}}
&\multicolumn{9}{l}{\tiny$\square$\normalsize~\it diagonal element $H_{mp,ff}$ contributions}\\[0.1cm]
$f$:&1&2&\dots&L&L+1&L+2&\dots&L(L+1)/2-1&L(L+1)/2\\
$\epsilon_ic^+_ic_i$:&$\epsilon_1+\epsilon_1$&$\epsilon_1+\epsilon_2$&$\dots$&$\epsilon_1+\epsilon_L$&$\epsilon_2+\epsilon_2$&$\epsilon_2+\epsilon_3$&$\dots$&$\epsilon_{L-1}+\epsilon_L$&$\epsilon_L+\epsilon_L$\\
$U_{i,j}c_i^+c_j^+c_ic_j$:&$U^\pm_L(0)$&$U^\pm_L(1)$&$\dots$&$U^\pm_L(L-1)$&$U^\pm_L(0)$&$U^\pm_L(1)$&$\dots$&$U^\pm_L(1)$&$U^\pm_L(0)$\\[0.3cm]
&\multicolumn{9}{l}{\tiny$\square$\normalsize~\it first off--diagonal elements $H_{mp,ff+1}/J$}\\[0.1cm]
$Jc^+_{i+1}c_i$:&$\sqrt{2}$&$1$&$\dots$&$0$&$\sqrt{2}$&$1$&$\dots$&$0$&$-$\\[0.3cm]
&\multicolumn{9}{l}{\tiny$\square$\normalsize~\it remaining off--diagonal elements $H_{mp,ff'}$}\\[0.1cm]
$Jc^+_{i+1}c_i$:&\multicolumn{9}{l}{all remaining off--diagonal elements have magnitude
$J$. In the $j$th off--diagonal \brc{$j\geq1$} there are $j$}\\[0.1cm]
&\multicolumn{9}{l}{non-zero elements: $H_{mp,ff+j}$ with}\\[0.1cm]
&\multicolumn{9}{c}{$f=L-\sum^j_{r=1}r=L-j(j+1)/2~,\quad\dots~,\quad L-j(j+1)/2-(j-1)\quad.$}
\end{tabular}
\end{center}
\end{widetext}

\end{document}

%% file: FIG1.tex
\begingroup
  \fontfamily{cmss}%
  \selectfont
  \makeatletter
  \providecommand\color[2][]{%
    \GenericError{(gnuplot) \space\space\space\@spaces}{%
      Package color not loaded in conjunction with
      terminal option `colourtext'%
    }{See the gnuplot documentation for explanation.%
    }{Either use 'blacktext' in gnuplot or load the package
      color.sty in LaTeX.}%
    \renewcommand\color[2][]{}%
  }%
  \providecommand\includegraphics[2][]{%
    \GenericError{(gnuplot) \space\space\space\@spaces}{%
      Package graphicx or graphics not loaded%
    }{See the gnuplot documentation for explanation.%
    }{The gnuplot epslatex terminal needs graphicx.sty or graphics.sty.}%
    \renewcommand\includegraphics[2][]{}%
  }%
  \providecommand\rotatebox[2]{#2}%
  \@ifundefined{ifGPcolor}{%
    \newif\ifGPcolor
    \GPcolortrue
  }{}%
  \@ifundefined{ifGPblacktext}{%
    \newif\ifGPblacktext
    \GPblacktextfalse
  }{}%
  \let\gplgaddtomacro\g@addto@macro
  \gdef\gplbacktext{}%
  \gdef\gplfronttext{}%
  \makeatother
  \ifGPblacktext
    \def\colorrgb#1{}%
    \def\colorgray#1{}%
  \else
    \ifGPcolor
      \def\colorrgb#1{\color[rgb]{#1}}%
      \def\colorgray#1{\color[gray]{#1}}%
      \expandafter\def\csname LTw\endcsname{\color{white}}%
      \expandafter\def\csname LTb\endcsname{\color{black}}%
      \expandafter\def\csname LTa\endcsname{\color{black}}%
      \expandafter\def\csname LT0\endcsname{\color[rgb]{1,0,0}}%
      \expandafter\def\csname LT1\endcsname{\color[rgb]{0,1,0}}%
      \expandafter\def\csname LT2\endcsname{\color[rgb]{0,0,1}}%
      \expandafter\def\csname LT3\endcsname{\color[rgb]{1,0,1}}%
      \expandafter\def\csname LT4\endcsname{\color[rgb]{0,1,1}}%
      \expandafter\def\csname LT5\endcsname{\color[rgb]{1,1,0}}%
      \expandafter\def\csname LT6\endcsname{\color[rgb]{0,0,0}}%
      \expandafter\def\csname LT7\endcsname{\color[rgb]{1,0.3,0}}%
      \expandafter\def\csname LT8\endcsname{\color[rgb]{0.5,0.5,0.5}}%
    \else
      \def\colorrgb#1{\color{black}}%
      \def\colorgray#1{\color[gray]{#1}}%
      \expandafter\def\csname LTw\endcsname{\color{white}}%
      \expandafter\def\csname LTb\endcsname{\color{black}}%
      \expandafter\def\csname LTa\endcsname{\color{black}}%
      \expandafter\def\csname LT0\endcsname{\color{black}}%
      \expandafter\def\csname LT1\endcsname{\color{black}}%
      \expandafter\def\csname LT2\endcsname{\color{black}}%
      \expandafter\def\csname LT3\endcsname{\color{black}}%
      \expandafter\def\csname LT4\endcsname{\color{black}}%
      \expandafter\def\csname LT5\endcsname{\color{black}}%
      \expandafter\def\csname LT6\endcsname{\color{black}}%
      \expandafter\def\csname LT7\endcsname{\color{black}}%
      \expandafter\def\csname LT8\endcsname{\color{black}}%
    \fi
  \fi
  \setlength{\unitlength}{0.0500bp}%
  \begin{picture}(7200.00,5760.00)%
    \gplgaddtomacro\gplbacktext{%
      \csname LTb\endcsname%
      \put(3599,5187){\makebox(0,0){\strut{}LL at $\kappa=1$}}%
      \put(220,220){\makebox(0,0)[l]{\strut{}}}%
    }%
    \gplgaddtomacro\gplfronttext{%
      \csname LTb\endcsname%
      \put(1656,558){\makebox(0,0){\strut{}-0.4}}%
      \put(2628,558){\makebox(0,0){\strut{}-0.2}}%
      \put(3600,558){\makebox(0,0){\strut{} 0}}%
      \put(4572,558){\makebox(0,0){\strut{} 0.2}}%
      \put(5544,558){\makebox(0,0){\strut{} 0.4}}%
      \put(3600,228){\makebox(0,0){\strut{}energy $e$}}%
      \put(998,844){\makebox(0,0)[r]{\strut{} 0}}%
      \put(998,1514){\makebox(0,0)[r]{\strut{} 0.5}}%
      \put(998,2185){\makebox(0,0)[r]{\strut{} 1}}%
      \put(998,2855){\makebox(0,0)[r]{\strut{} 1.5}}%
      \put(998,3525){\makebox(0,0)[r]{\strut{} 2}}%
      \put(998,4196){\makebox(0,0)[r]{\strut{} 2.5}}%
      \put(998,4866){\makebox(0,0)[r]{\strut{} 3}}%
      \put(404,2855){\rotatebox{90}{\makebox(0,0){\strut{}correlation parameter $\alpha$}}}%
      \put(6527,843){\makebox(0,0)[l]{\strut{} 0}}%
      \put(6527,1848){\makebox(0,0)[l]{\strut{} 0.2}}%
      \put(6527,2854){\makebox(0,0)[l]{\strut{} 0.4}}%
      \put(6527,3860){\makebox(0,0)[l]{\strut{} 0.6}}%
      \put(6527,4866){\makebox(0,0)[l]{\strut{} 0.8}}%
      \put(7121,2854){\rotatebox{90}{\makebox(0,0){\strut{}}}}%
    }%
    \gplgaddtomacro\gplbacktext{%
      \put(1580,1622){\makebox(0,0){\relsize{-2} 0.2}}%
      \put(1580,2125){\makebox(0,0){\relsize{-2} 0.4}}%
      \put(1580,2628){\makebox(0,0){\relsize{-2} 0.6}}%
      \put(2142,1020){\makebox(0,0){\relsize{-2} 1.5}}%
      \put(2520,1020){\makebox(0,0){\strut{}}}%
      \put(2897,1020){\makebox(0,0){\relsize{-2} 2.5}}%
      \put(1390,2016){\rotatebox{90}{\makebox(0,0){\relsize{-1}LL($e=0$)}}}%
      \put(3494,2016){\rotatebox{90}{\makebox(0,0){\strut{}}}}%
      \put(2519,997){\makebox(0,0){\relsize{-1}$\alpha$}}%
      \put(2519,2769){\makebox(0,0){\strut{}}}%
      \put(-876,953){\makebox(0,0)[l]{\strut{}}}%
    }%
    \gplgaddtomacro\gplfronttext{%
      \csname LTb\endcsname%
      \put(4460,2141){\makebox(0,0){\relsize{-1}L$=$1500\quad\quad\quad}}%
      \csname LTb\endcsname%
      \put(4460,1921){\makebox(0,0){\relsize{-1}L$=$2500\quad\quad\quad}}%
      \csname LTb\endcsname%
      \put(4460,1701){\makebox(0,0){\relsize{-1}L$=$3500\quad\quad\quad}}%
    }%
    \put(0,0){\includegraphics{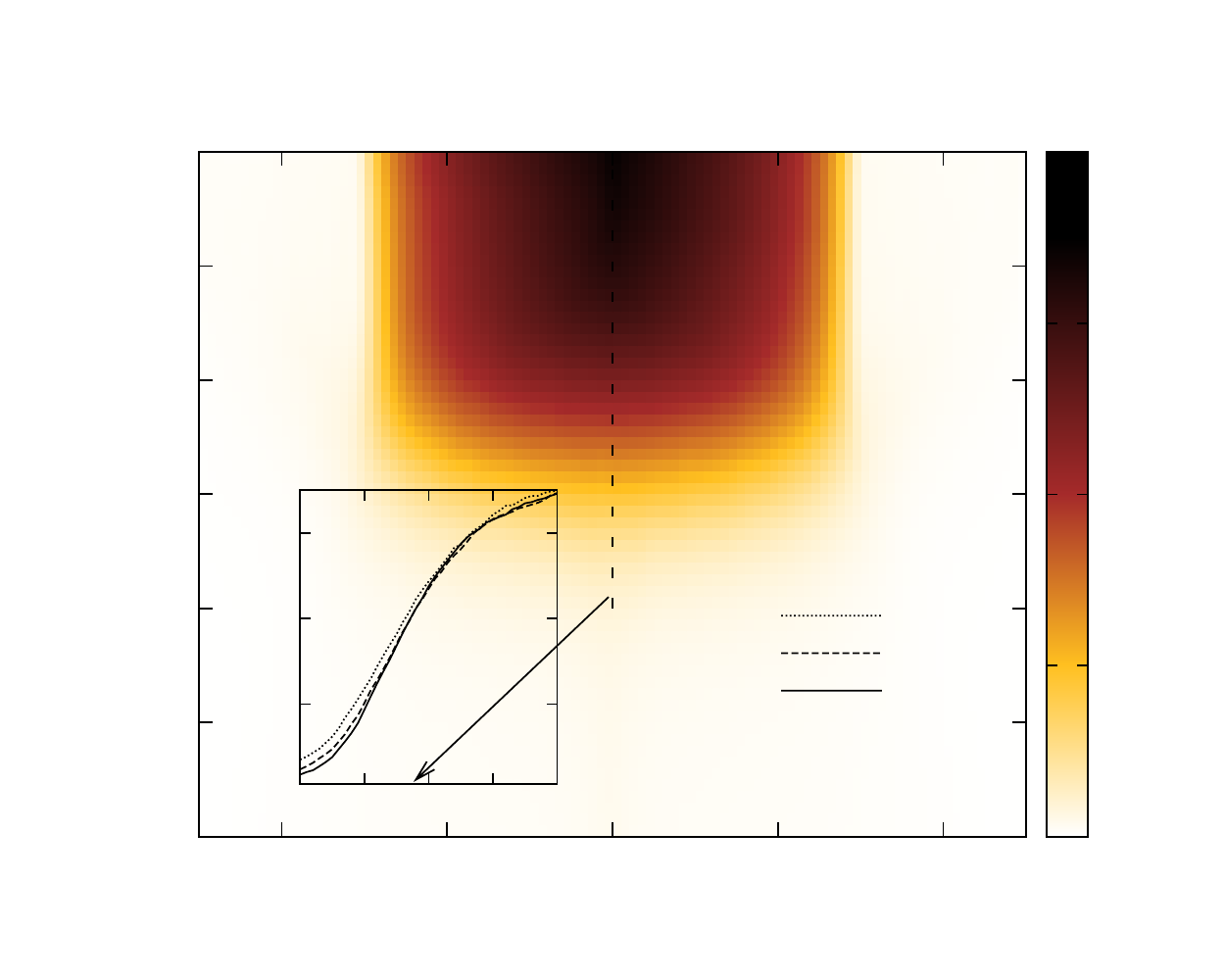}}%
\gplbacktext    \gplfronttext
  \end{picture}%
\endgroup

%% file: FIG2.tex
\begingroup
  \fontfamily{cmss}%
  \selectfont
  \makeatletter
  \providecommand\color[2][]{%
    \GenericError{(gnuplot) \space\space\space\@spaces}{%
      Package color not loaded in conjunction with
      terminal option `colourtext'%
    }{See the gnuplot documentation for explanation.%
    }{Either use 'blacktext' in gnuplot or load the package
      color.sty in LaTeX.}%
    \renewcommand\color[2][]{}%
  }%
  \providecommand\includegraphics[2][]{%
    \GenericError{(gnuplot) \space\space\space\@spaces}{%
      Package graphicx or graphics not loaded%
    }{See the gnuplot documentation for explanation.%
    }{The gnuplot epslatex terminal needs graphicx.sty or graphics.sty.}%
    \renewcommand\includegraphics[2][]{}%
  }%
  \providecommand\rotatebox[2]{#2}%
  \@ifundefined{ifGPcolor}{%
    \newif\ifGPcolor
    \GPcolortrue
  }{}%
  \@ifundefined{ifGPblacktext}{%
    \newif\ifGPblacktext
    \GPblacktextfalse
  }{}%
  \let\gplgaddtomacro\g@addto@macro
  \gdef\gplbacktext{}%
  \gdef\gplfronttext{}%
  \makeatother
  \ifGPblacktext
    \def\colorrgb#1{}%
    \def\colorgray#1{}%
  \else
    \ifGPcolor
      \def\colorrgb#1{\color[rgb]{#1}}%
      \def\colorgray#1{\color[gray]{#1}}%
      \expandafter\def\csname LTw\endcsname{\color{white}}%
      \expandafter\def\csname LTb\endcsname{\color{black}}%
      \expandafter\def\csname LTa\endcsname{\color{black}}%
      \expandafter\def\csname LT0\endcsname{\color[rgb]{1,0,0}}%
      \expandafter\def\csname LT1\endcsname{\color[rgb]{0,1,0}}%
      \expandafter\def\csname LT2\endcsname{\color[rgb]{0,0,1}}%
      \expandafter\def\csname LT3\endcsname{\color[rgb]{1,0,1}}%
      \expandafter\def\csname LT4\endcsname{\color[rgb]{0,1,1}}%
      \expandafter\def\csname LT5\endcsname{\color[rgb]{1,1,0}}%
      \expandafter\def\csname LT6\endcsname{\color[rgb]{0,0,0}}%
      \expandafter\def\csname LT7\endcsname{\color[rgb]{1,0.3,0}}%
      \expandafter\def\csname LT8\endcsname{\color[rgb]{0.5,0.5,0.5}}%
    \else
      \def\colorrgb#1{\color{black}}%
      \def\colorgray#1{\color[gray]{#1}}%
      \expandafter\def\csname LTw\endcsname{\color{white}}%
      \expandafter\def\csname LTb\endcsname{\color{black}}%
      \expandafter\def\csname LTa\endcsname{\color{black}}%
      \expandafter\def\csname LT0\endcsname{\color{black}}%
      \expandafter\def\csname LT1\endcsname{\color{black}}%
      \expandafter\def\csname LT2\endcsname{\color{black}}%
      \expandafter\def\csname LT3\endcsname{\color{black}}%
      \expandafter\def\csname LT4\endcsname{\color{black}}%
      \expandafter\def\csname LT5\endcsname{\color{black}}%
      \expandafter\def\csname LT6\endcsname{\color{black}}%
      \expandafter\def\csname LT7\endcsname{\color{black}}%
      \expandafter\def\csname LT8\endcsname{\color{black}}%
    \fi
  \fi
  \setlength{\unitlength}{0.0500bp}%
  \begin{picture}(7200.00,5760.00)%
    \gplgaddtomacro\gplbacktext{%
      \csname LTb\endcsname%
      \put(3599,5309){\makebox(0,0){\strut{}NSD distribution $P(\text{NSD})$ at $\kappa=1$}}%
      \put(220,220){\makebox(0,0)[l]{\strut{}}}%
    }%
    \gplgaddtomacro\gplfronttext{%
      \csname LTb\endcsname%
      \put(1170,667){\makebox(0,0){\strut{} 0}}%
      \put(1980,667){\makebox(0,0){\strut{} 0.2}}%
      \put(2790,667){\makebox(0,0){\strut{} 0.4}}%
      \put(3600,667){\makebox(0,0){\strut{} 0.6}}%
      \put(4410,667){\makebox(0,0){\strut{} 0.8}}%
      \put(5220,667){\makebox(0,0){\strut{} 1}}%
      \put(6030,667){\makebox(0,0){\strut{} 1.2}}%
      \put(3600,337){\makebox(0,0){\strut{}NSD}}%
      \put(998,953){\makebox(0,0)[r]{\strut{} 0}}%
      \put(998,1618){\makebox(0,0)[r]{\strut{} 0.5}}%
      \put(998,2283){\makebox(0,0)[r]{\strut{} 1}}%
      \put(998,2947){\makebox(0,0)[r]{\strut{} 1.5}}%
      \put(998,3611){\makebox(0,0)[r]{\strut{} 2}}%
      \put(998,4276){\makebox(0,0)[r]{\strut{} 2.5}}%
      \put(998,4941){\makebox(0,0)[r]{\strut{} 3}}%
      \put(404,2947){\rotatebox{90}{\makebox(0,0){\strut{}correlation parameter $\alpha$}}}%
      \put(6527,952){\makebox(0,0)[l]{\strut{}0}}%
      \put(6527,2281){\makebox(0,0)[l]{\strut{}0.1}}%
      \put(6527,3611){\makebox(0,0)[l]{\strut{}0.2}}%
      \put(6527,4941){\makebox(0,0)[l]{\strut{}0.3}}%
      \put(6989,2946){\rotatebox{90}{\makebox(0,0){\strut{}}}}%
    }%
    \gplgaddtomacro\gplbacktext{%
      \put(1709,3283){\rotatebox{90}{\makebox(0,0){\relsize{-2} 1$\cdot$10$^{\text{-4}}$}}}%
      \put(1709,3859){\rotatebox{90}{\makebox(0,0){\relsize{-2} 2$\cdot$10$^{\text{-4}}$}}}%
      \put(1709,4434){\rotatebox{90}{\makebox(0,0){\relsize{-2} 3$\cdot$10$^{\text{-4}}$}}}%
      \put(1871,2863){\makebox(0,0){\relsize{-2} 0}}%
      \put(2579,2863){\makebox(0,0){\relsize{-2} 1}}%
      \put(3287,2863){\makebox(0,0){\relsize{-2} 2}}%
      \put(3996,2863){\makebox(0,0){\relsize{-2} 3}}%
      \put(1497,3858){\rotatebox{90}{\makebox(0,0){\relsize{-1}$\overline{\text{NSD}}$}}}%
      \put(4250,3858){\rotatebox{90}{\makebox(0,0){\strut{}}}}%
      \put(2951,2840){\makebox(0,0){\relsize{-1}$\alpha$}}%
      \put(2951,4611){\makebox(0,0){\strut{}}}%
      \put(1211,2796){\makebox(0,0)[l]{\strut{}}}%
    }%
    \gplgaddtomacro\gplfronttext{%
      \put(3353,3864){\makebox(0,0){\relsize{-1}L$=$1500}}%
      \put(3353,3644){\makebox(0,0){\relsize{-1}L$=$2500}}%
      \put(3353,3424){\makebox(0,0){\relsize{-1}L$=$3500}}%
    }%
    \put(0,0){\includegraphics{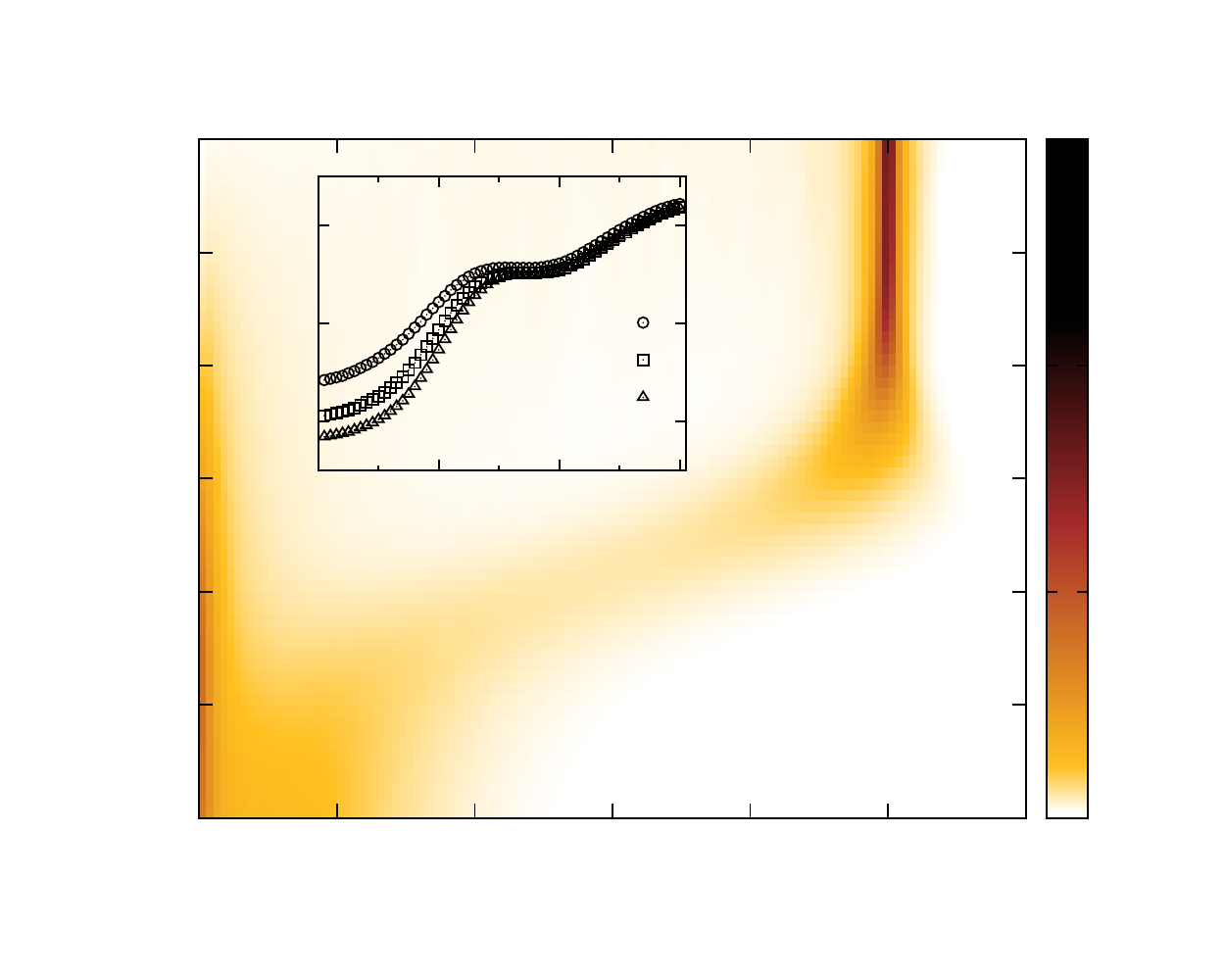}}%
\gplbacktext    \gplfronttext
  \end{picture}%
\endgroup

%% file: FIG3a.tex
\begingroup
  \fontfamily{cmss}%
  \selectfont
  \makeatletter
  \providecommand\color[2][]{%
    \GenericError{(gnuplot) \space\space\space\@spaces}{%
      Package color not loaded in conjunction with
      terminal option `colourtext'%
    }{See the gnuplot documentation for explanation.%
    }{Either use 'blacktext' in gnuplot or load the package
      color.sty in LaTeX.}%
    \renewcommand\color[2][]{}%
  }%
  \providecommand\includegraphics[2][]{%
    \GenericError{(gnuplot) \space\space\space\@spaces}{%
      Package graphicx or graphics not loaded%
    }{See the gnuplot documentation for explanation.%
    }{The gnuplot epslatex terminal needs graphicx.sty or graphics.sty.}%
    \renewcommand\includegraphics[2][]{}%
  }%
  \providecommand\rotatebox[2]{#2}%
  \@ifundefined{ifGPcolor}{%
    \newif\ifGPcolor
    \GPcolortrue
  }{}%
  \@ifundefined{ifGPblacktext}{%
    \newif\ifGPblacktext
    \GPblacktextfalse
  }{}%
  \let\gplgaddtomacro\g@addto@macro
  \gdef\gplbacktext{}%
  \gdef\gplfronttext{}%
  \makeatother
  \ifGPblacktext
    \def\colorrgb#1{}%
    \def\colorgray#1{}%
  \else
    \ifGPcolor
      \def\colorrgb#1{\color[rgb]{#1}}%
      \def\colorgray#1{\color[gray]{#1}}%
      \expandafter\def\csname LTw\endcsname{\color{white}}%
      \expandafter\def\csname LTb\endcsname{\color{black}}%
      \expandafter\def\csname LTa\endcsname{\color{black}}%
      \expandafter\def\csname LT0\endcsname{\color[rgb]{1,0,0}}%
      \expandafter\def\csname LT1\endcsname{\color[rgb]{0,1,0}}%
      \expandafter\def\csname LT2\endcsname{\color[rgb]{0,0,1}}%
      \expandafter\def\csname LT3\endcsname{\color[rgb]{1,0,1}}%
      \expandafter\def\csname LT4\endcsname{\color[rgb]{0,1,1}}%
      \expandafter\def\csname LT5\endcsname{\color[rgb]{1,1,0}}%
      \expandafter\def\csname LT6\endcsname{\color[rgb]{0,0,0}}%
      \expandafter\def\csname LT7\endcsname{\color[rgb]{1,0.3,0}}%
      \expandafter\def\csname LT8\endcsname{\color[rgb]{0.5,0.5,0.5}}%
    \else
      \def\colorrgb#1{\color{black}}%
      \def\colorgray#1{\color[gray]{#1}}%
      \expandafter\def\csname LTw\endcsname{\color{white}}%
      \expandafter\def\csname LTb\endcsname{\color{black}}%
      \expandafter\def\csname LTa\endcsname{\color{black}}%
      \expandafter\def\csname LT0\endcsname{\color{black}}%
      \expandafter\def\csname LT1\endcsname{\color{black}}%
      \expandafter\def\csname LT2\endcsname{\color{black}}%
      \expandafter\def\csname LT3\endcsname{\color{black}}%
      \expandafter\def\csname LT4\endcsname{\color{black}}%
      \expandafter\def\csname LT5\endcsname{\color{black}}%
      \expandafter\def\csname LT6\endcsname{\color{black}}%
      \expandafter\def\csname LT7\endcsname{\color{black}}%
      \expandafter\def\csname LT8\endcsname{\color{black}}%
    \fi
  \fi
  \setlength{\unitlength}{0.0500bp}%
  \begin{picture}(7200.00,4320.00)%
    \gplgaddtomacro\gplbacktext{%
      \csname LTb\endcsname%
      \put(3599,4023){\makebox(0,0){\strut{}IPR \brc{top} and DOS \brc{bottom} at $\kappa=1$}}%
      \put(220,220){\makebox(0,0)[l]{\strut{}}}%
    }%
    \gplgaddtomacro\gplfronttext{%
      \csname LTb\endcsname%
      \put(1666,-208){\makebox(0,0){\strut{}}}%
      \put(2658,-208){\makebox(0,0){\strut{}}}%
      \put(3649,-208){\makebox(0,0){\strut{}}}%
      \put(4641,-208){\makebox(0,0){\strut{}}}%
      \put(5634,-208){\makebox(0,0){\strut{}}}%
      \put(3600,-538){\makebox(0,0){\strut{}}}%
      \put(998,78){\makebox(0,0)[r]{\strut{} 0}}%
      \put(998,662){\makebox(0,0)[r]{\strut{} 0.5}}%
      \put(998,1246){\makebox(0,0)[r]{\strut{} 1}}%
      \put(998,1830){\makebox(0,0)[r]{\strut{} 1.5}}%
      \put(998,2414){\makebox(0,0)[r]{\strut{} 2}}%
      \put(998,2998){\makebox(0,0)[r]{\strut{} 2.5}}%
      \put(998,3582){\makebox(0,0)[r]{\strut{} 3}}%
      \put(404,1830){\rotatebox{90}{\makebox(0,0){\strut{}correlation parameter $\alpha$}}}%
      \put(6527,77){\makebox(0,0)[l]{\strut{} 0}}%
      \put(6527,1245){\makebox(0,0)[l]{\strut{} 0.04}}%
      \put(6527,2413){\makebox(0,0)[l]{\strut{} 0.08}}%
      \put(6527,3582){\makebox(0,0)[l]{\strut{} 0.12}}%
      \put(7253,1829){\rotatebox{90}{\makebox(0,0){\strut{}}}}%
    }%
    \gplgaddtomacro\gplbacktext{%
      \put(2250,1322){\rotatebox{90}{\makebox(0,0){\relsize{-2} 3$\cdot$10$^{\text{-6}}$}}}%
      \put(2250,2199){\rotatebox{90}{\makebox(0,0){\relsize{-2} 6$\cdot$10$^{\text{-6}}$}}}%
      \put(2250,3077){\rotatebox{90}{\makebox(0,0){\relsize{-2} 9$\cdot$10$^{\text{-6}}$}}}%
      \put(2412,429){\makebox(0,0){\relsize{-2} 0}}%
      \put(3356,429){\makebox(0,0){\relsize{-2} 1}}%
      \put(4300,429){\makebox(0,0){\relsize{-2} 2}}%
      \put(5244,429){\makebox(0,0){\relsize{-2} 3}}%
      \put(2038,1965){\rotatebox{90}{\makebox(0,0){\relsize{-1}$\overline{\text{IPR}}$}}}%
      \put(5510,1965){\rotatebox{90}{\makebox(0,0){\strut{}}}}%
      \put(3851,406){\makebox(0,0){\relsize{-1}$\alpha$}}%
      \put(3851,3258){\makebox(0,0){\strut{}}}%
      \put(1752,362){\makebox(0,0)[l]{\strut{}}}%
    }%
    \gplgaddtomacro\gplfronttext{%
      \put(4483,2967){\makebox(0,0){\relsize{-1}L$=$1500}}%
      \put(4483,2747){\makebox(0,0){\relsize{-1}L$=$2500}}%
      \put(4483,2527){\makebox(0,0){\relsize{-1}L$=$3500}}%
    }%
    \put(0,0){\includegraphics{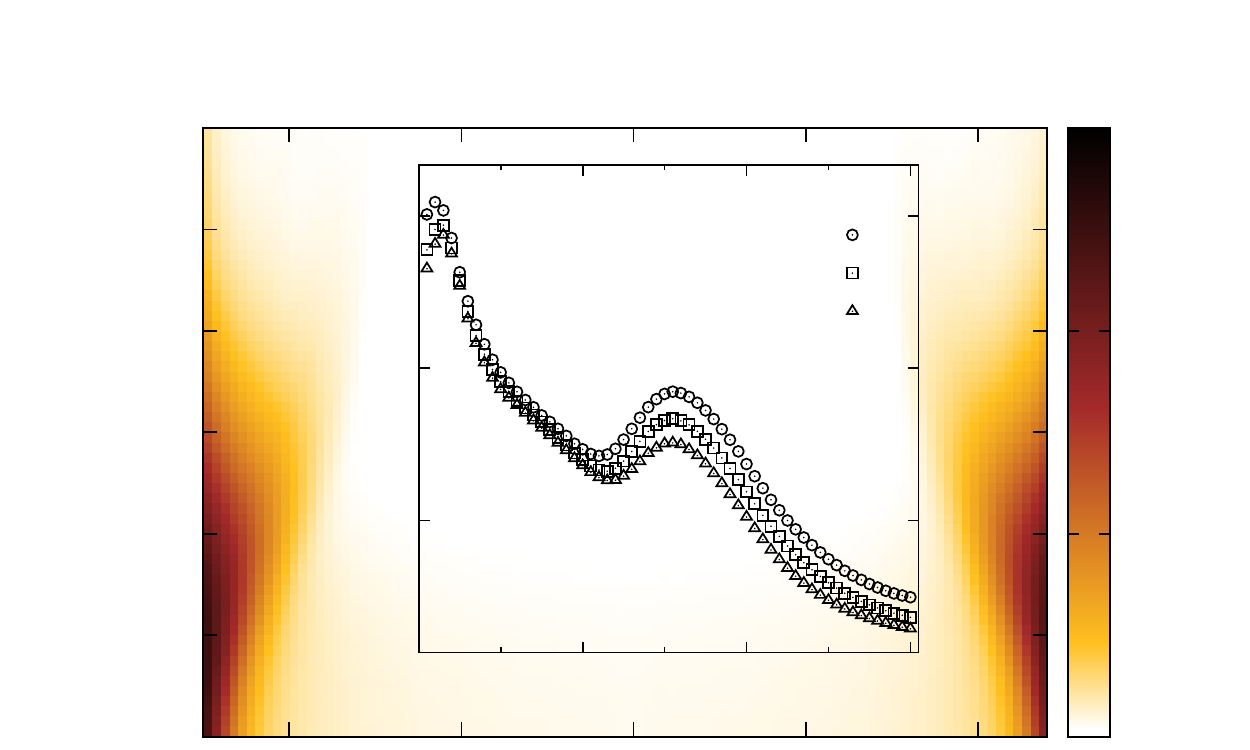}}%
\gplbacktext    \gplfronttext
  \end{picture}%
\endgroup

%% file: FIG3b.tex
\begingroup
  \fontfamily{cmss}%
  \selectfont
  \makeatletter
  \providecommand\color[2][]{%
    \GenericError{(gnuplot) \space\space\space\@spaces}{%
      Package color not loaded in conjunction with
      terminal option `colourtext'%
    }{See the gnuplot documentation for explanation.%
    }{Either use 'blacktext' in gnuplot or load the package
      color.sty in LaTeX.}%
    \renewcommand\color[2][]{}%
  }%
  \providecommand\includegraphics[2][]{%
    \GenericError{(gnuplot) \space\space\space\@spaces}{%
      Package graphicx or graphics not loaded%
    }{See the gnuplot documentation for explanation.%
    }{The gnuplot epslatex terminal needs graphicx.sty or graphics.sty.}%
    \renewcommand\includegraphics[2][]{}%
  }%
  \providecommand\rotatebox[2]{#2}%
  \@ifundefined{ifGPcolor}{%
    \newif\ifGPcolor
    \GPcolortrue
  }{}%
  \@ifundefined{ifGPblacktext}{%
    \newif\ifGPblacktext
    \GPblacktextfalse
  }{}%
  \let\gplgaddtomacro\g@addto@macro
  \gdef\gplbacktext{}%
  \gdef\gplfronttext{}%
  \makeatother
  \ifGPblacktext
    \def\colorrgb#1{}%
    \def\colorgray#1{}%
  \else
    \ifGPcolor
      \def\colorrgb#1{\color[rgb]{#1}}%
      \def\colorgray#1{\color[gray]{#1}}%
      \expandafter\def\csname LTw\endcsname{\color{white}}%
      \expandafter\def\csname LTb\endcsname{\color{black}}%
      \expandafter\def\csname LTa\endcsname{\color{black}}%
      \expandafter\def\csname LT0\endcsname{\color[rgb]{1,0,0}}%
      \expandafter\def\csname LT1\endcsname{\color[rgb]{0,1,0}}%
      \expandafter\def\csname LT2\endcsname{\color[rgb]{0,0,1}}%
      \expandafter\def\csname LT3\endcsname{\color[rgb]{1,0,1}}%
      \expandafter\def\csname LT4\endcsname{\color[rgb]{0,1,1}}%
      \expandafter\def\csname LT5\endcsname{\color[rgb]{1,1,0}}%
      \expandafter\def\csname LT6\endcsname{\color[rgb]{0,0,0}}%
      \expandafter\def\csname LT7\endcsname{\color[rgb]{1,0.3,0}}%
      \expandafter\def\csname LT8\endcsname{\color[rgb]{0.5,0.5,0.5}}%
    \else
      \def\colorrgb#1{\color{black}}%
      \def\colorgray#1{\color[gray]{#1}}%
      \expandafter\def\csname LTw\endcsname{\color{white}}%
      \expandafter\def\csname LTb\endcsname{\color{black}}%
      \expandafter\def\csname LTa\endcsname{\color{black}}%
      \expandafter\def\csname LT0\endcsname{\color{black}}%
      \expandafter\def\csname LT1\endcsname{\color{black}}%
      \expandafter\def\csname LT2\endcsname{\color{black}}%
      \expandafter\def\csname LT3\endcsname{\color{black}}%
      \expandafter\def\csname LT4\endcsname{\color{black}}%
      \expandafter\def\csname LT5\endcsname{\color{black}}%
      \expandafter\def\csname LT6\endcsname{\color{black}}%
      \expandafter\def\csname LT7\endcsname{\color{black}}%
      \expandafter\def\csname LT8\endcsname{\color{black}}%
    \fi
  \fi
  \setlength{\unitlength}{0.0500bp}%
  \begin{picture}(7200.00,3600.00)%
    \gplgaddtomacro\gplbacktext{%
      \csname LTb\endcsname%
      \put(3599,3826){\makebox(0,0){\strut{}}}%
      \put(220,220){\makebox(0,0)[l]{\strut{}}}%
    }%
    \gplgaddtomacro\gplfronttext{%
      \csname LTb\endcsname%
      \put(1621,392){\makebox(0,0){\strut{}-0.4}}%
      \put(2623,392){\makebox(0,0){\strut{}-0.2}}%
      \put(3625,392){\makebox(0,0){\strut{} 0}}%
      \put(4627,392){\makebox(0,0){\strut{} 0.2}}%
      \put(5630,392){\makebox(0,0){\strut{} 0.4}}%
      \put(3600,62){\makebox(0,0){\strut{}energy $e$}}%
      \put(998,678){\makebox(0,0)[r]{\strut{} 0}}%
      \put(998,1148){\makebox(0,0)[r]{\strut{} 0.5}}%
      \put(998,1618){\makebox(0,0)[r]{\strut{} 1}}%
      \put(998,2087){\makebox(0,0)[r]{\strut{} 1.5}}%
      \put(998,2556){\makebox(0,0)[r]{\strut{} 2}}%
      \put(998,3026){\makebox(0,0)[r]{\strut{} 2.5}}%
      \put(998,3496){\makebox(0,0)[r]{\strut{} 3}}%
      \put(404,2087){\rotatebox{90}{\makebox(0,0){\strut{}correlation parameter $\alpha$}}}%
      \put(6527,677){\makebox(0,0)[l]{\strut{} 0}}%
      \put(6527,1616){\makebox(0,0)[l]{\strut{} 0.01}}%
      \put(6527,2556){\makebox(0,0)[l]{\strut{} 0.02}}%
      \put(6527,3496){\makebox(0,0)[l]{\strut{} 0.03}}%
      \put(7253,2086){\rotatebox{90}{\makebox(0,0){\strut{}}}}%
    }%
    \gplbacktext
    \put(0,0){\includegraphics{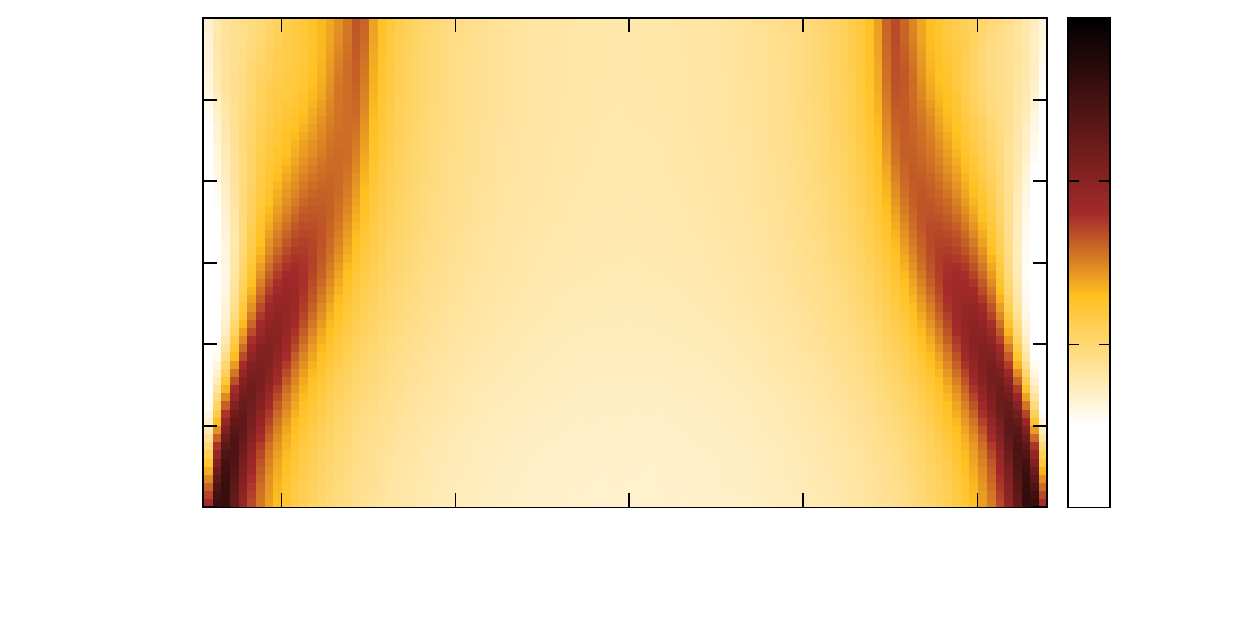}}%
    \gplfronttext
  \end{picture}%
\endgroup

%% file: FIG4.tex
\begingroup
  \fontfamily{cmss}%
  \selectfont
  \makeatletter
  \providecommand\color[2][]{%
    \GenericError{(gnuplot) \space\space\space\@spaces}{%
      Package color not loaded in conjunction with
      terminal option `colourtext'%
    }{See the gnuplot documentation for explanation.%
    }{Either use 'blacktext' in gnuplot or load the package
      color.sty in LaTeX.}%
    \renewcommand\color[2][]{}%
  }%
  \providecommand\includegraphics[2][]{%
    \GenericError{(gnuplot) \space\space\space\@spaces}{%
      Package graphicx or graphics not loaded%
    }{See the gnuplot documentation for explanation.%
    }{The gnuplot epslatex terminal needs graphicx.sty or graphics.sty.}%
    \renewcommand\includegraphics[2][]{}%
  }%
  \providecommand\rotatebox[2]{#2}%
  \@ifundefined{ifGPcolor}{%
    \newif\ifGPcolor
    \GPcolortrue
  }{}%
  \@ifundefined{ifGPblacktext}{%
    \newif\ifGPblacktext
    \GPblacktextfalse
  }{}%
  \let\gplgaddtomacro\g@addto@macro
  \gdef\gplbacktext{}%
  \gdef\gplfronttext{}%
  \makeatother
  \ifGPblacktext
    \def\colorrgb#1{}%
    \def\colorgray#1{}%
  \else
    \ifGPcolor
      \def\colorrgb#1{\color[rgb]{#1}}%
      \def\colorgray#1{\color[gray]{#1}}%
      \expandafter\def\csname LTw\endcsname{\color{white}}%
      \expandafter\def\csname LTb\endcsname{\color{black}}%
      \expandafter\def\csname LTa\endcsname{\color{black}}%
      \expandafter\def\csname LT0\endcsname{\color[rgb]{1,0,0}}%
      \expandafter\def\csname LT1\endcsname{\color[rgb]{0,1,0}}%
      \expandafter\def\csname LT2\endcsname{\color[rgb]{0,0,1}}%
      \expandafter\def\csname LT3\endcsname{\color[rgb]{1,0,1}}%
      \expandafter\def\csname LT4\endcsname{\color[rgb]{0,1,1}}%
      \expandafter\def\csname LT5\endcsname{\color[rgb]{1,1,0}}%
      \expandafter\def\csname LT6\endcsname{\color[rgb]{0,0,0}}%
      \expandafter\def\csname LT7\endcsname{\color[rgb]{1,0.3,0}}%
      \expandafter\def\csname LT8\endcsname{\color[rgb]{0.5,0.5,0.5}}%
    \else
      \def\colorrgb#1{\color{black}}%
      \def\colorgray#1{\color[gray]{#1}}%
      \expandafter\def\csname LTw\endcsname{\color{white}}%
      \expandafter\def\csname LTb\endcsname{\color{black}}%
      \expandafter\def\csname LTa\endcsname{\color{black}}%
      \expandafter\def\csname LT0\endcsname{\color{black}}%
      \expandafter\def\csname LT1\endcsname{\color{black}}%
      \expandafter\def\csname LT2\endcsname{\color{black}}%
      \expandafter\def\csname LT3\endcsname{\color{black}}%
      \expandafter\def\csname LT4\endcsname{\color{black}}%
      \expandafter\def\csname LT5\endcsname{\color{black}}%
      \expandafter\def\csname LT6\endcsname{\color{black}}%
      \expandafter\def\csname LT7\endcsname{\color{black}}%
      \expandafter\def\csname LT8\endcsname{\color{black}}%
    \fi
  \fi
  \setlength{\unitlength}{0.0500bp}%
  \begin{picture}(7200.00,6480.00)%
    \gplgaddtomacro\gplbacktext{%
      \csname LTb\endcsname%
      \put(1188,660){\makebox(0,0)[r]{\strut{}1$\cdot$10$^{\text{-2}}$}}%
      \put(1188,2423){\makebox(0,0)[r]{\strut{}1$\cdot$10$^{\text{-1}}$}}%
      \put(1188,4186){\makebox(0,0)[r]{\strut{}1$\cdot$10$^{\text{0}}$}}%
      \put(1188,5948){\makebox(0,0)[r]{\strut{}1$\cdot$10$^{\text{1}}$}}%
      \put(1320,440){\makebox(0,0){\strut{}0}}%
      \put(2790,440){\makebox(0,0){\strut{}1}}%
      \put(4260,440){\makebox(0,0){\strut{}2}}%
      \put(5729,440){\makebox(0,0){\strut{}3}}%
      \put(7199,440){\makebox(0,0){\strut{}4}}%
      \put(22,3569){\rotatebox{-270}{\makebox(0,0){\strut{}nearest neighbor distribution $P(s)$}}}%
      \put(4259,110){\makebox(0,0){\strut{}nearest neighbor spacing $s$}}%
    }%
    \gplgaddtomacro\gplfronttext{%
      \csname LTb\endcsname%
      \put(5987,6023){\makebox(0,0)[r]{\strut{}$P_\text{P}$}}%
      \csname LTb\endcsname%
      \put(5987,5551){\makebox(0,0)[r]{\strut{}$P_\text{WD}$}}%
      \csname LTb\endcsname%
      \put(5987,5079){\makebox(0,0){$\alpha=$ 0.0\quad\quad}}%
      \csname LTb\endcsname%
      \put(5987,4607){\makebox(0,0){$\alpha=$ 0.5\quad\quad}}%
      \csname LTb\endcsname%
      \put(5987,4135){\makebox(0,0){$\alpha=$ 1.0\quad\quad}}%
      \csname LTb\endcsname%
      \put(5987,3663){\makebox(0,0){$\alpha=$ 1.5\quad\quad}}%
      \csname LTb\endcsname%
      \put(5987,3191){\makebox(0,0){$\alpha=$ 2.0\quad\quad}}%
      \csname LTb\endcsname%
      \put(5987,2719){\makebox(0,0){$\alpha=$ 2.5\quad\quad}}%
    }%
    \gplbacktext
    \put(0,0){\includegraphics{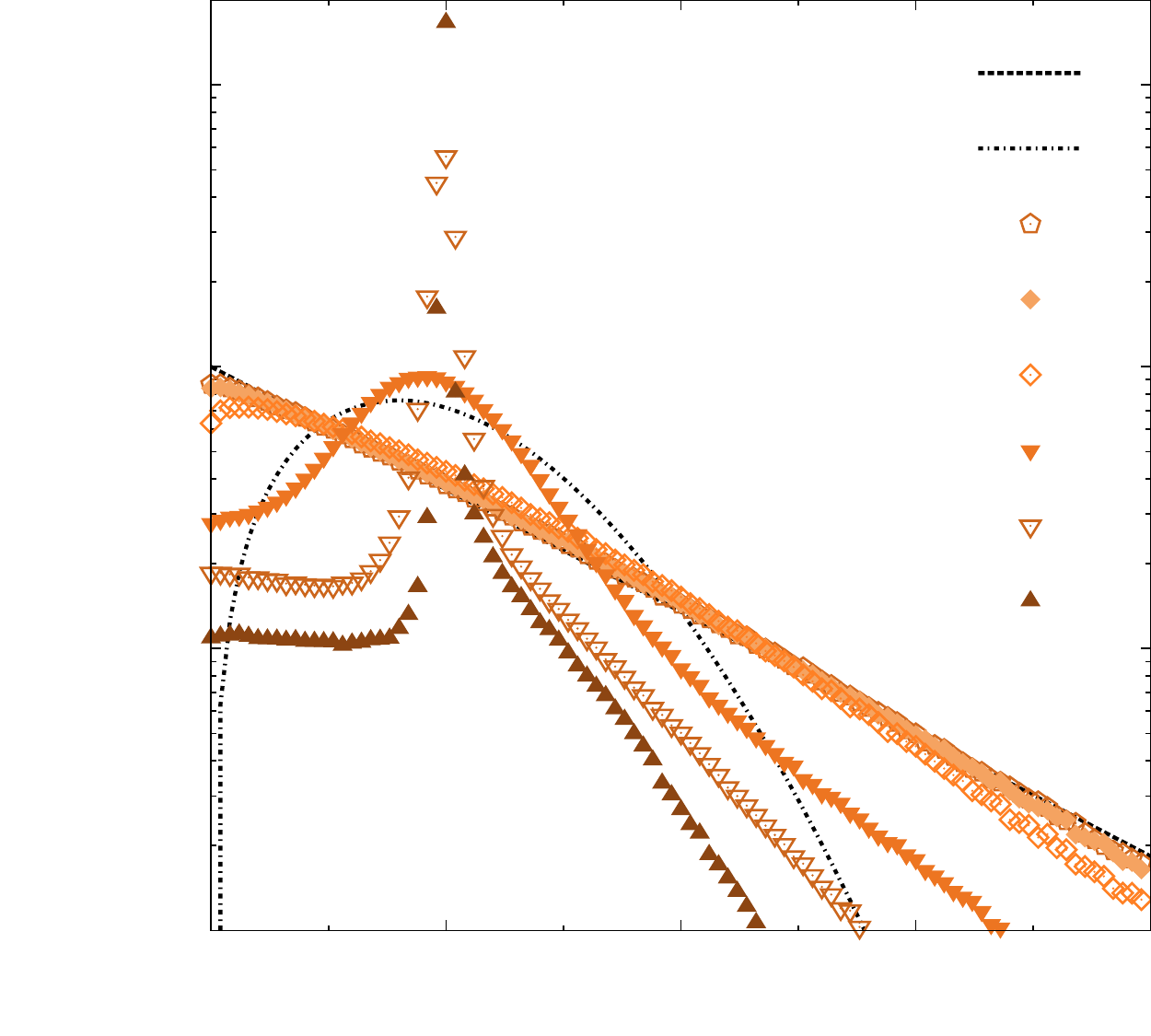}}%
    \gplfronttext
  \end{picture}%
\endgroup

%% file: FIG5.tex
\begingroup
  \fontfamily{cmss}%
  \selectfont
  \makeatletter
  \providecommand\color[2][]{%
    \GenericError{(gnuplot) \space\space\space\@spaces}{%
      Package color not loaded in conjunction with
      terminal option `colourtext'%
    }{See the gnuplot documentation for explanation.%
    }{Either use 'blacktext' in gnuplot or load the package
      color.sty in LaTeX.}%
    \renewcommand\color[2][]{}%
  }%
  \providecommand\includegraphics[2][]{%
    \GenericError{(gnuplot) \space\space\space\@spaces}{%
      Package graphicx or graphics not loaded%
    }{See the gnuplot documentation for explanation.%
    }{The gnuplot epslatex terminal needs graphicx.sty or graphics.sty.}%
    \renewcommand\includegraphics[2][]{}%
  }%
  \providecommand\rotatebox[2]{#2}%
  \@ifundefined{ifGPcolor}{%
    \newif\ifGPcolor
    \GPcolortrue
  }{}%
  \@ifundefined{ifGPblacktext}{%
    \newif\ifGPblacktext
    \GPblacktextfalse
  }{}%
  \let\gplgaddtomacro\g@addto@macro
  \gdef\gplbacktext{}%
  \gdef\gplfronttext{}%
  \makeatother
  \ifGPblacktext
    \def\colorrgb#1{}%
    \def\colorgray#1{}%
  \else
    \ifGPcolor
      \def\colorrgb#1{\color[rgb]{#1}}%
      \def\colorgray#1{\color[gray]{#1}}%
      \expandafter\def\csname LTw\endcsname{\color{white}}%
      \expandafter\def\csname LTb\endcsname{\color{black}}%
      \expandafter\def\csname LTa\endcsname{\color{black}}%
      \expandafter\def\csname LT0\endcsname{\color[rgb]{1,0,0}}%
      \expandafter\def\csname LT1\endcsname{\color[rgb]{0,1,0}}%
      \expandafter\def\csname LT2\endcsname{\color[rgb]{0,0,1}}%
      \expandafter\def\csname LT3\endcsname{\color[rgb]{1,0,1}}%
      \expandafter\def\csname LT4\endcsname{\color[rgb]{0,1,1}}%
      \expandafter\def\csname LT5\endcsname{\color[rgb]{1,1,0}}%
      \expandafter\def\csname LT6\endcsname{\color[rgb]{0,0,0}}%
      \expandafter\def\csname LT7\endcsname{\color[rgb]{1,0.3,0}}%
      \expandafter\def\csname LT8\endcsname{\color[rgb]{0.5,0.5,0.5}}%
    \else
      \def\colorrgb#1{\color{black}}%
      \def\colorgray#1{\color[gray]{#1}}%
      \expandafter\def\csname LTw\endcsname{\color{white}}%
      \expandafter\def\csname LTb\endcsname{\color{black}}%
      \expandafter\def\csname LTa\endcsname{\color{black}}%
      \expandafter\def\csname LT0\endcsname{\color{black}}%
      \expandafter\def\csname LT1\endcsname{\color{black}}%
      \expandafter\def\csname LT2\endcsname{\color{black}}%
      \expandafter\def\csname LT3\endcsname{\color{black}}%
      \expandafter\def\csname LT4\endcsname{\color{black}}%
      \expandafter\def\csname LT5\endcsname{\color{black}}%
      \expandafter\def\csname LT6\endcsname{\color{black}}%
      \expandafter\def\csname LT7\endcsname{\color{black}}%
      \expandafter\def\csname LT8\endcsname{\color{black}}%
    \fi
  \fi
  \setlength{\unitlength}{0.0500bp}%
  \begin{picture}(7200.00,8640.00)%
    \gplgaddtomacro\gplbacktext{%
      \csname LTb\endcsname%
      \put(3599,8365){\makebox(0,0){\strut{}$L_\text{NND}$}}%
      \put(220,220){\makebox(0,0)[l]{\strut{}}}%
      \put(1277,7393){\makebox(0,0)[l]{\strut{}EXTENDED}}%
      \put(1277,6830){\makebox(0,0)[l]{\strut{}due to Ref.[\onlinecite{Shim04PhysRevB}]}}%
    }%
    \gplgaddtomacro\gplfronttext{%
      \csname LTb\endcsname%
      \put(1980,491){\makebox(0,0){\strut{} 2}}%
      \put(3060,491){\makebox(0,0){\strut{} 4}}%
      \put(4140,491){\makebox(0,0){\strut{} 6}}%
      \put(5220,491){\makebox(0,0){\strut{} 8}}%
      \put(3600,161){\makebox(0,0){\strut{}disorder strength $\kappa$}}%
      \put(998,777){\makebox(0,0)[r]{\strut{} 0}}%
      \put(998,2185){\makebox(0,0)[r]{\strut{} 0.5}}%
      \put(998,3593){\makebox(0,0)[r]{\strut{} 1}}%
      \put(998,4999){\makebox(0,0)[r]{\strut{} 1.5}}%
      \put(998,6407){\makebox(0,0)[r]{\strut{} 2}}%
      \put(998,7815){\makebox(0,0)[r]{\strut{} 2.5}}%
      \put(404,4296){\rotatebox{90}{\makebox(0,0){\strut{}correlation parameter $\alpha$}}}%
      \put(6527,1705){\makebox(0,0)[l]{\strut{}10$^\text{-4}$}}%
      \put(6527,3033){\makebox(0,0)[l]{\strut{}10$^\text{-3}$}}%
      \put(6527,4361){\makebox(0,0)[l]{\strut{}10$^\text{-2}$}}%
      \put(6527,5690){\makebox(0,0)[l]{\strut{}10$^\text{-1}$}}%
      \put(6527,7018){\makebox(0,0)[l]{\strut{}10$^\text{0}$}}%
      \put(8441,4295){\rotatebox{90}{\makebox(0,0){\strut{}}}}%
    }%
    \gplgaddtomacro\gplbacktext{%
      \csname LT0\endcsname%
      \put(3798,1571){\rotatebox{90}{\makebox(0,0)[r]{\strut{}\textcolor{white}{\bf\relsize{-2}10$^{\text{-2}}$}}}}%
      \put(3798,2416){\rotatebox{90}{\makebox(0,0)[r]{\strut{}\textcolor{white}{\bf\relsize{-2}10$^{\text{-1}}$}}}}%
      \put(3798,3261){\rotatebox{90}{\makebox(0,0)[r]{\strut{}\textcolor{white}{\bf\relsize{-2}10$^{\text{0}}$}}}}%
      \put(3798,4105){\rotatebox{90}{\makebox(0,0)[r]{\strut{}\textcolor{white}{\bf\relsize{-2}10$^{\text{1}}$}}}}%
      \put(4383,1164){\makebox(0,0){\strut{}\textcolor{white}{\bf\relsize{-2} 5000}}}%
      \put(4912,1164){\makebox(0,0){\strut{}}}%
      \put(5442,1164){\makebox(0,0){\strut{}\textcolor{white}{\bf\relsize{-2} 15000}}}%
      \csname LTb\endcsname%
      \put(3322,2764){\rotatebox{90}{\makebox(0,0){\strut{}\textcolor{white}{\bf\relsize{-1}$\Delta[L^2_\text{NND}]^2$}}}}%
      \put(5978,2764){\rotatebox{90}{\makebox(0,0){\strut{}}}}%
      \put(4859,966){\makebox(0,0){\strut{}\textcolor{white}{\bf\relsize{-1}$L$}}}%
      \put(4859,4122){\makebox(0,0){\strut{}}}%
      \put(3300,922){\makebox(0,0)[l]{\strut{}}}%
    }%
    \gplgaddtomacro\gplfronttext{%
    }%
    \put(0,0){\includegraphics{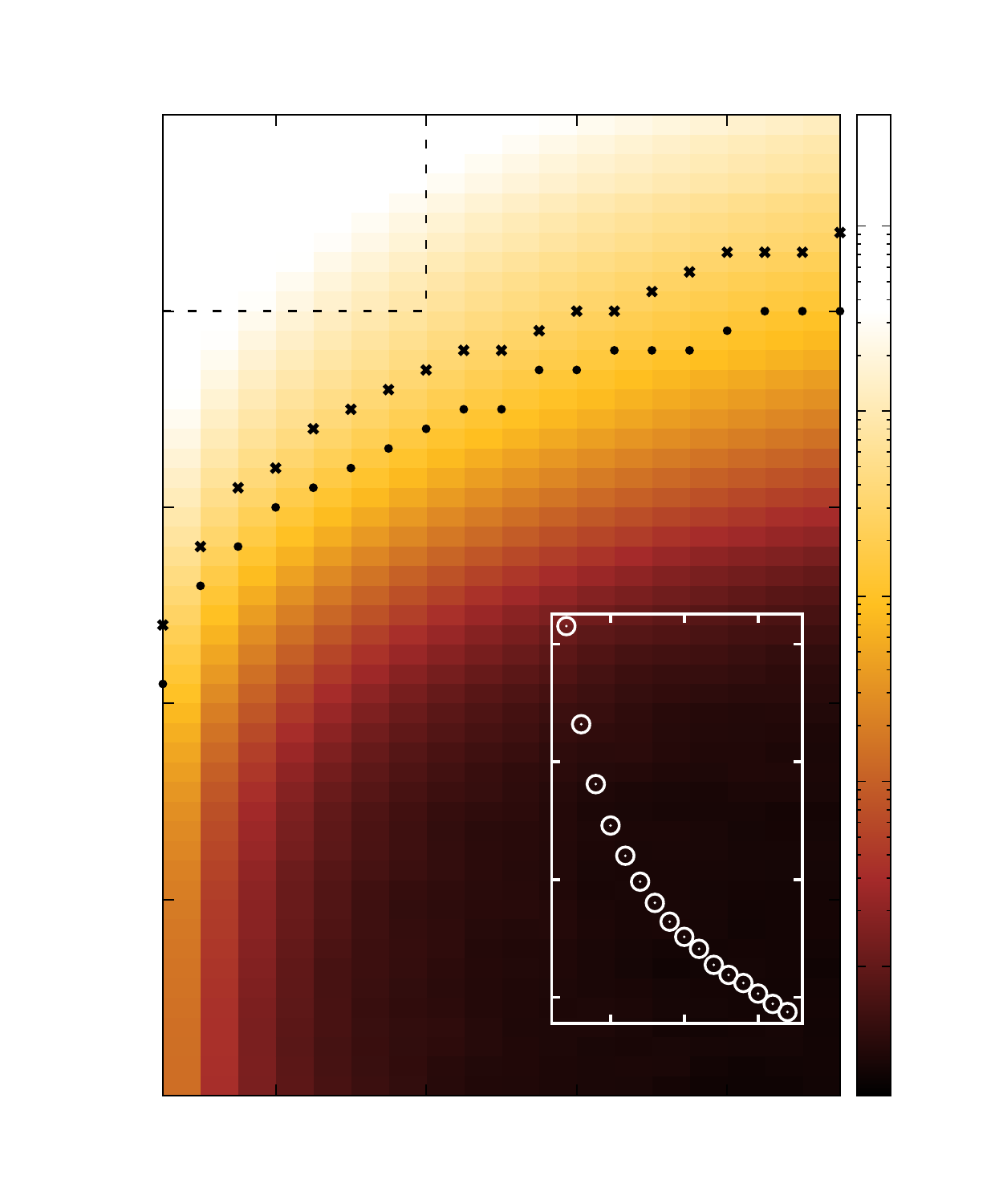}}%
\gplbacktext    \gplfronttext
  \end{picture}%
\endgroup

%% file: FIG6.tex
\begingroup
  \fontfamily{cmss}%
  \selectfont
  \makeatletter
  \providecommand\color[2][]{%
    \GenericError{(gnuplot) \space\space\space\@spaces}{%
      Package color not loaded in conjunction with
      terminal option `colourtext'%
    }{See the gnuplot documentation for explanation.%
    }{Either use 'blacktext' in gnuplot or load the package
      color.sty in LaTeX.}%
    \renewcommand\color[2][]{}%
  }%
  \providecommand\includegraphics[2][]{%
    \GenericError{(gnuplot) \space\space\space\@spaces}{%
      Package graphicx or graphics not loaded%
    }{See the gnuplot documentation for explanation.%
    }{The gnuplot epslatex terminal needs graphicx.sty or graphics.sty.}%
    \renewcommand\includegraphics[2][]{}%
  }%
  \providecommand\rotatebox[2]{#2}%
  \@ifundefined{ifGPcolor}{%
    \newif\ifGPcolor
    \GPcolortrue
  }{}%
  \@ifundefined{ifGPblacktext}{%
    \newif\ifGPblacktext
    \GPblacktextfalse
  }{}%
  \let\gplgaddtomacro\g@addto@macro
  \gdef\gplbacktext{}%
  \gdef\gplfronttext{}%
  \makeatother
  \ifGPblacktext
    \def\colorrgb#1{}%
    \def\colorgray#1{}%
  \else
    \ifGPcolor
      \def\colorrgb#1{\color[rgb]{#1}}%
      \def\colorgray#1{\color[gray]{#1}}%
      \expandafter\def\csname LTw\endcsname{\color{white}}%
      \expandafter\def\csname LTb\endcsname{\color{black}}%
      \expandafter\def\csname LTa\endcsname{\color{black}}%
      \expandafter\def\csname LT0\endcsname{\color[rgb]{1,0,0}}%
      \expandafter\def\csname LT1\endcsname{\color[rgb]{0,1,0}}%
      \expandafter\def\csname LT2\endcsname{\color[rgb]{0,0,1}}%
      \expandafter\def\csname LT3\endcsname{\color[rgb]{1,0,1}}%
      \expandafter\def\csname LT4\endcsname{\color[rgb]{0,1,1}}%
      \expandafter\def\csname LT5\endcsname{\color[rgb]{1,1,0}}%
      \expandafter\def\csname LT6\endcsname{\color[rgb]{0,0,0}}%
      \expandafter\def\csname LT7\endcsname{\color[rgb]{1,0.3,0}}%
      \expandafter\def\csname LT8\endcsname{\color[rgb]{0.5,0.5,0.5}}%
    \else
      \def\colorrgb#1{\color{black}}%
      \def\colorgray#1{\color[gray]{#1}}%
      \expandafter\def\csname LTw\endcsname{\color{white}}%
      \expandafter\def\csname LTb\endcsname{\color{black}}%
      \expandafter\def\csname LTa\endcsname{\color{black}}%
      \expandafter\def\csname LT0\endcsname{\color{black}}%
      \expandafter\def\csname LT1\endcsname{\color{black}}%
      \expandafter\def\csname LT2\endcsname{\color{black}}%
      \expandafter\def\csname LT3\endcsname{\color{black}}%
      \expandafter\def\csname LT4\endcsname{\color{black}}%
      \expandafter\def\csname LT5\endcsname{\color{black}}%
      \expandafter\def\csname LT6\endcsname{\color{black}}%
      \expandafter\def\csname LT7\endcsname{\color{black}}%
      \expandafter\def\csname LT8\endcsname{\color{black}}%
    \fi
  \fi
  \setlength{\unitlength}{0.0500bp}%
  \begin{picture}(21600.00,7200.00)%
    \gplgaddtomacro\gplbacktext{%
      \csname LTb\endcsname%
      \put(1595,720){\makebox(0,0)[r]{\strut{} 0}}%
      \put(1595,2340){\makebox(0,0)[r]{\strut{} 0.2}}%
      \put(1595,3960){\makebox(0,0)[r]{\strut{} 0.4}}%
      \put(1595,5579){\makebox(0,0)[r]{\strut{} 0.6}}%
      \put(1595,7199){\makebox(0,0)[r]{\strut{} 0.8}}%
      \put(1595,500){\makebox(0,0){\strut{} 0}}%
      \put(3215,500){\makebox(0,0){\strut{} 1}}%
      \put(4835,500){\makebox(0,0){\strut{} 2}}%
      \put(6455,500){\makebox(0,0){\strut{} 3}}%
      \put(8075,500){\makebox(0,0){\strut{} 4}}%
      \put(429,3959){\rotatebox{-270}{\makebox(0,0){\strut{}$P(s)$}}}%
      \put(4835,170){\makebox(0,0){\strut{}$s$}}%
      \put(2537,1530){\makebox(0,0)[l]{\strut{}$\kappa =$ 0.5}}%
    }%
    \gplgaddtomacro\gplfronttext{%
      \csname LTb\endcsname%
      \put(7220,6821){\makebox(0,0)[r]{\strut{}$\alpha_\text{min} =$ 1.05}}%
      \csname LTb\endcsname%
      \put(7220,6191){\makebox(0,0)[r]{\strut{}$\alpha=\alpha_\text{min} -$ 0.1}}%
      \csname LTb\endcsname%
      \put(7220,5561){\makebox(0,0)[r]{\strut{}$\alpha=\alpha_\text{min} +$ 0.1}}%
      \csname LTb\endcsname%
      \put(7220,4931){\makebox(0,0)[r]{\strut{}$P_\text{SP}$}}%
    }%
    \gplgaddtomacro\gplbacktext{%
      \csname LTb\endcsname%
      \put(8291,720){\makebox(0,0)[r]{\strut{}}}%
      \put(8291,2340){\makebox(0,0)[r]{\strut{}}}%
      \put(8291,3960){\makebox(0,0)[r]{\strut{}}}%
      \put(8291,5579){\makebox(0,0)[r]{\strut{}}}%
      \put(8291,7199){\makebox(0,0)[r]{\strut{}}}%
      \put(8357,500){\makebox(0,0){\strut{} 0}}%
      \put(9977,500){\makebox(0,0){\strut{} 1}}%
      \put(11597,500){\makebox(0,0){\strut{} 2}}%
      \put(13217,500){\makebox(0,0){\strut{} 3}}%
      \put(14837,500){\makebox(0,0){\strut{} 4}}%
      \put(11531,170){\makebox(0,0){\strut{}$s$}}%
      \put(8731,2097){\makebox(0,0)[l]{\strut{}$\alpha_\text{min} =$ 1.85}}%
      \put(9233,1530){\makebox(0,0)[l]{\strut{}$\kappa =$ 5.5}}%
    }%
    \gplgaddtomacro\gplfronttext{%
    }%
    \gplgaddtomacro\gplbacktext{%
      \csname LTb\endcsname%
      \put(14987,720){\makebox(0,0)[r]{\strut{}}}%
      \put(14987,2340){\makebox(0,0)[r]{\strut{}}}%
      \put(14987,3960){\makebox(0,0)[r]{\strut{}}}%
      \put(14987,5579){\makebox(0,0)[r]{\strut{}}}%
      \put(14987,7199){\makebox(0,0)[r]{\strut{}}}%
      \put(15119,500){\makebox(0,0){\strut{} 0}}%
      \put(16736,500){\makebox(0,0){\strut{} 1}}%
      \put(18353,500){\makebox(0,0){\strut{} 2}}%
      \put(19969,500){\makebox(0,0){\strut{} 3}}%
      \put(21586,500){\makebox(0,0){\strut{} 4}}%
      \put(18220,170){\makebox(0,0){\strut{}$s$}}%
      \put(15426,2097){\makebox(0,0)[l]{\strut{}$\alpha_\text{min} =$ 2.00}}%
      \put(15927,1530){\makebox(0,0)[l]{\strut{}$\kappa =$ 9.5}}%
    }%
    \gplgaddtomacro\gplfronttext{%
    }%
    \gplgaddtomacro\gplbacktext{%
      \csname LTb\endcsname%
      \put(18875,3239){\makebox(0,0)[r]{\strut{}}}%
      \put(18875,3839){\makebox(0,0)[r]{\strut{}}}%
      \put(18875,4439){\makebox(0,0)[r]{\strut{}}}%
      \put(18875,5039){\makebox(0,0)[r]{\strut{}}}%
      \put(18875,5639){\makebox(0,0)[r]{\strut{}}}%
      \put(18875,6239){\makebox(0,0)[r]{\strut{}}}%
      \put(18875,6839){\makebox(0,0)[r]{\strut{}}}%
      \put(19007,3019){\makebox(0,0){\relsize{-2}0}}%
      \put(19439,3019){\makebox(0,0){\relsize{-2}2}}%
      \put(19871,3019){\makebox(0,0){\relsize{-2}4}}%
      \put(20303,3019){\makebox(0,0){\relsize{-2}6}}%
      \put(20735,3019){\makebox(0,0){\relsize{-2}8}}%
      \put(21167,3019){\makebox(0,0){\relsize{-2}10}}%
      \put(18633,5039){\rotatebox{-270}{\makebox(0,0){\relsize{-2} deviation from $P_\text{SP}$ \brc{arb. units}}}}%
      \put(20087,2689){\makebox(0,0){\relsize{-2}disorder strength $\kappa$}}%
    }%
    \gplgaddtomacro\gplfronttext{%
    }%
    \gplbacktext
    \put(0,0){\includegraphics{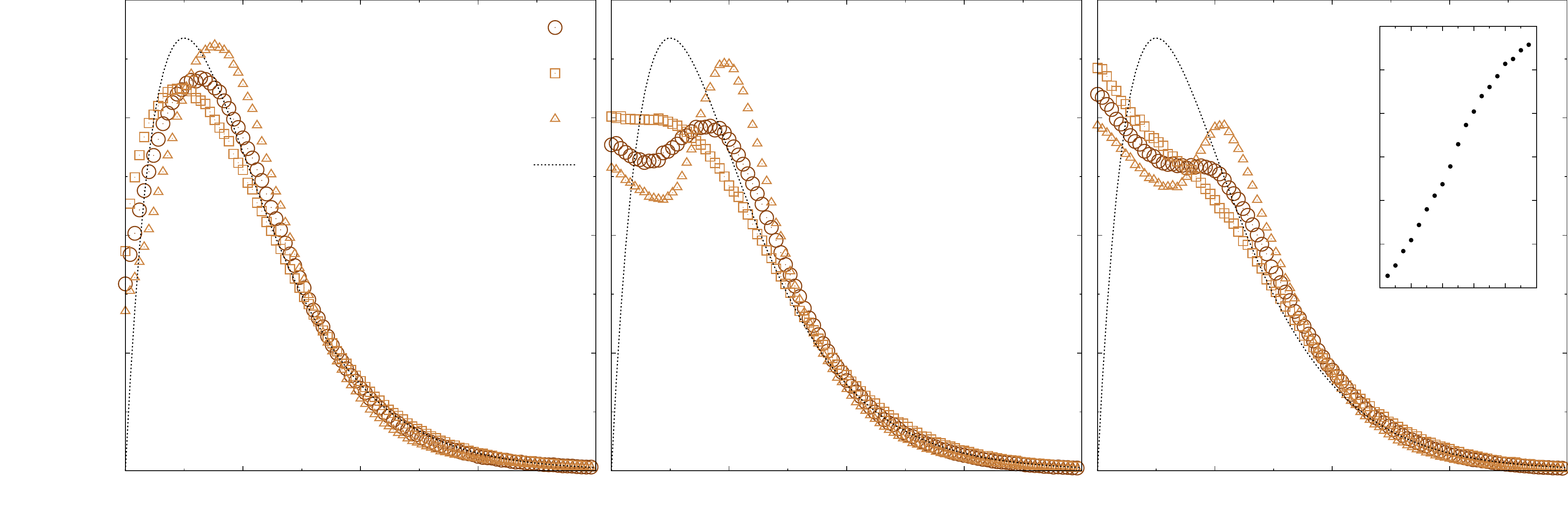}}%
    \gplfronttext
  \end{picture}%
\endgroup